\documentclass[letter]{sig-alternate-10pt} 

\usepackage{times}
\usepackage{amsmath}
\usepackage{graphicx}
\usepackage{url}
\usepackage[latin1]{inputenc}
\usepackage{lastpage}
\usepackage{xspace}
\usepackage{subfig}

\newcommand{\comment}[1]{}

\newcounter{packednmbr}
\newenvironment{packedenumerate}{\begin{list}{\thepackednmbr.}{\usecounter{packednmbr}\setlength{\itemsep}{0pt}\addtolength{\labelwidth}{-5pt}\setlength{\leftmargin}{\labelwidth}\setlength{\listparindent}{\parindent}\setlength{\parsep}{0pt}\setlength{\topsep}{3pt}}}{\end{list}}
\newenvironment{packeditemize}{\begin{list}{$\bullet$}{\setlength{\itemsep}{0pt}\addtolength{\labelwidth}{-5pt}\setlength{\leftmargin}{\labelwidth}\setlength{\listparindent}{\parindent}\setlength{\parsep}{0pt}\setlength{\topsep}{3pt}}}{\end{list}}

\newcommand{\tightcaption}[1]{\vspace{-0.3cm}\caption{#1}\vspace{-0.2cm}}

\newcommand{\ignore}[1]{}
\newcommand{\figref}[1]{Fig.~\ref{#1}}

\begin{document}
\pdfpagewidth=8.5in
\pdfpageheight=11in

\title{CARE: Content Aware Redundancy Elimination for Disaster Communications on Damaged Networks}

\author{
Udi Weinsberg$^1$ \and Athula Balachandran$^2$ \and Nina Taft$^1$ \and Gianluca Iannaccone$^3$ \and  Vyas Sekar$^4$ \and Srinivasan Seshan$^2$}

\maketitle

\pagenumbering{arabic}
\pagestyle{plain}

\begin{abstract}

During a disaster scenario, situational awareness information, such as
location, physical status and images of the surrounding area, is essential for
minimizing loss of life, injury, and property damage. Today's handhelds make it
easy for people to gather data from within the disaster area in many formats,
including text, images and video. Studies show that the extreme anxiety
induced  by disasters causes humans to create a substantial amount of
repetitive and  redundant content. Transporting this
content outside the disaster zone can be problematic when  the  network
infrastructure is disrupted  by the disaster.

This paper presents the design of a novel architecture called CARE
(Content-Aware Redundancy Elimination)  for better utilizing network resources
in disaster-affected regions.  Motivated by
measurement-driven insights on redundancy patterns found in real-world disaster area photos, 
we demonstrate that CARE
can detect the \emph{semantic} similarity between photos in the networking layer, thus 
reducing redundant transfers and improving buffer utilization. 
Using DTN simulations, we explore
the boundaries of the usefulness of deploying CARE on a damaged network, and
show that CARE can reduce packet delivery times and drops, and enables 20-40\%
more unique information to reach the rescue teams outside the disaster area
than when CARE is not deployed.

\end{abstract}

\section{Introduction}

{\renewcommand{\thefootnote}{\fnsymbol{footnote}}
\footnotetext[0]{\scriptsize
1) Technicolor, Palo-Alto Research Lab}
\footnotetext[0]{\scriptsize
2) Computer Science Department, Carnegie Mellon University}
\footnotetext[0]{\scriptsize
3) Red Bow Labs}
\footnotetext[0]{\scriptsize
4) Intel Labs}
\renewcommand{\thefootnote}{\arabic{footnote}}
}

During disaster events, such as fires, floods and earthquakes, situational
awareness (SA) information is critical for victims and rescue workers to make
well-informed decisions. SA information can include many things, such as
location information (of fires, floods, damaged homes), the status of the
disaster (e.g., fire temperature, flood water levels), resource information
(e.g., rescue, medical, water and food resources), and the status of
individuals (e.g., health monitoring, distress calls). It is essential to
gather data from inside a disaster area and deliver it to a service where it
can be aggregated, processed and effectively shared. Recently we have witnessed
an upsurge in the use of online social networks, such as Twitter
\cite{twitter-use-09} and Facebook \cite{facebook-use-08}, as well as the
emergence of new Web-based applications
\cite{ushahidi,people-finder,mashups,portal-iscram08,hackathon} that can both
assist victims and enable ordinary citizens to involve themselves in the
emergency response effort.  A key contributing trend to these
new applications is that everyday devices such as smartphones and laptops
enable ordinary citizens inside a disaster zone to gather critical information
in a variety of formats, including text, images, audio and video.




These emerging applications, and uses of online social networks for SA, 
assume that those inside the disaster zone can actually connect to
the Internet. All too often this is not the case because the usual
communication infrastructure can be compromised by the disaster \cite{allman,
google-haiti-report, katrina, nyu-report}. Both Louisiana and Mississippi
suffered significant loss of Internet access after hurricane Katrina;  with
some towns remaining unreachable for prolonged periods of time \cite{katrina}.
Reports of the Haiti earthquake reveal that Haiti's telecommunications
infrastructure was  destroyed, leaving the country with spotty mobile
connections, no Internet connectivity and little power in the first days after
the earthquake \cite{nethope}.  Moreover it took nearly one month to recover to
50\% of its original capacity \cite{google-haiti-report}. Studies of disasters
show that telecommunications often fail during disasters as a result of
destruction of components,  infrastructure, and network congestion
\cite{nyu-report}.  Thus, a key  challenge in supporting these emerging
applications is to manage the flow of SA data, coming from inside a disaster
zone, when connectivity to the Internet may be disrupted or intermittent for
extended periods of time.



At the same time, studies  show  that in response to extreme anxiety caused by
disasters, humans tend to want to communicate continuously,  resulting  in
repetitive information~\cite{siteseeing}. When victims send ``I'm OK"
messages, or a photo of their burning house, repeatedly to everyone they know,
a lot of redundant content is created and transmitted.
 A user under  duress  is unlikely
 to carefully cull out such duplicate content;
  instead she would send all the photos to her friends, family, and
 online SA service.  Similarly, a crowd in front of an exploding building can
result in different people generating the same content.  The  ease with
which users can gather SA content today and share it with their friends only
exacerbates the potential for redundant content. This natural behavior
aggravates the problem of limited connectivity, and leads to congestion and
inefficient use of shared communication, storage, and power resources.

In this paper, we address the problem of enabling SA applications to work
properly in the face of network disruptions and congestion due to content
redundancy. We consider scenarios in which people inside the disaster zone can
exchange information they collect via ad hoc networking. Opportunities to
upload this content to SA applications on the Internet arise either when people
or rescue vehicles have an opportunity to leave the disaster zone and reach a
place where they can connect to the Internet, or when connectivity inside the
disaster zone is temporarily restored. We thus propose to use the Delay
Tolerant Networking (DTN) stack on mobile phones and laptops because DTN
enables this carry-store-forward paradigm.

\comment
{We handle content redundancy through the use of semantic aware traffic
reduction. In this study we focus on image data and use
techniques from the field of computer vision to identify redundant content in
photos. Much progress in the field of computer vision in the last decade has
produced practical algorithms for object recognition and scene recognition.
Because mobile phones have evolved into powerful image and video processing
devices, equipped with hardware-accelerated graphics, the computational cost of
running such algorithms on handheld devices has diminished
\cite{sift-on-mobile}.  In the near future, image similarity detection
algorithms will become a basic component of handheld devices, and this opens
the door to using such algorithms for traffic processing tasks. Once content
redundancy is identified, then options for resource management open up, such as
dropping redundant photos, placing them in the back of a queue, or potentially
merging a set of photos into a smaller aggregated data item.
{\bf Vyas: this last para seems out of context}
}

Beyond the  vision of articulating the challenges and opportunities
that arise in this class of emerging networking applications, this paper
makes the following contributions:

\begin{packeditemize}
\item We demonstrate the semantic redundancy in photo datasets from real
disasters. We do so by a careful manual labeling process in discussion with
experts who have acted as first-responders in actual disasters (Section~\ref{sec:image_redundancy}).

\item We propose to use computer vision algorithms in the forwarding path to identify similar images, and thus describe an architecture that augments DTN forwarding with content-aware reduction algorithms. Our architecture allows redundancy to be detected both on a single device and across different devices (e.g., two people photographing the same scene) (Section~\ref{sec:arch}).

\item We design and implement a hybrid approach for detecting similar
images that uses measurement-driven insights to  combine three state-of-art
computer vision algorithms  to balance the tradeoffs between
accuracy and computational cost (Section~\ref{sec:algo}).

\item Using a DTN simulator~\cite{dtn}, we explore the benefits that
an architecture like CARE can provide under a range of operating scenarios
(Section~\ref{sec:eval}).  We find that CARE can improve the amount of unique
information delivered outside the disaster area by 5--40\% depending upon the
conditions. Moreover, CARE  delivers this information with up to 40\% reduced
latency. If this saves even a few lives, it would be a huge win.

\end{packeditemize}

\section{Redundancy in Data from \\ Disaster-affected Regions}
\label{sec:image_redundancy}


In this section, we analyze two real-world image datasets generated by people
inside a disaster zone and quantify the extent and nature of redundancy
found in these datasets.  The results  motivate the need for semantic or content-aware
techniques in the network layer for identifying redundant content in contrast to
prior byte- or data-level  techniques.

\begin{packedenumerate}

\item {\em San Diego fire (SDfire)}. This dataset contains 84 pictures taken by a professional
photographer who wandered around one of the affected towns both during the fire
event in 2007 and afterwards \cite{karl}.  The pictures depict a variety of scenes including
burning homes, damaged homes and cars, firefighters, policemen, etc.  This
dataset serves as an example of the type of  data and  redundancy, that could
be generated by a single individual.

\item  {\em Haiti earthquake}.  This dataset contains 415 pictures taken during and after
the Haiti
earthquake in January 2010 by the volunteers of an organization called Team Rubicon \cite{TeamRubicon}.
Team Rubicon sends
small teams, of roughly 10 people that are primarily medical staff, into
disaster zones.
The photos cover a wide range of subjects
including wounded people, damaged buildings, rescue vehicles (jeeps,
helicopters), victims trapped under rubble, the medical staff, the service
stations, street scenes, famous buildings, and crowd formations of victims.
\end{packedenumerate}

\subsection{Identifying and Quantifying Redundancy}

In both  datasets, just perusing the data suggested that there were
several pairs of similar images. We also saw instances of 3 photos, or even 4
photos that were similar. To objectively evaluate the extent
 of redundancy or how automated techniques perform, we need a
 notion of ground truth. Thus, we decided to manually label the data ourselves
 and built a web-based tool to facilitate the labeling procedure.
Our tool lets the labeler view two images at a time and allows her to rate the
similarity of the two images on a scale from 0 to 5;  0 being  not similar
 and 5 denoting extremely similar (typical
Likert scale~\cite{likert}).

The notion of ``similar" or ``redundant" (which we use interchangeably) is  difficult to define precisely.
Some cases are straightforward, e.g., two photos of the same scene where they differ in focus, luminance or resolution. 
However, consider a more complex scenario. Suppose
photos A and B show the same person in different backgrounds (say 
different places on the same block), and this person appears safe. If the goal
of the situational awareness application is merely to find people, or to report
whether or not people are ok, then these two images can be considered
redundant. However if the goal  is to track a person's movements, then the two
photos are not redundant. Similarly, if the background contains a burning house
in one image but not in the other, and the application evaluates where
help is needed, then these two images are not redundant. It becomes
clear that the notion of similarity is subjective, and more
importantly, ultimately depends upon the ``intended use" of the data. Recall
that our focus is on situational awareness applications, and their goal
 includes things like enabling victims to find out where help is
located, or where road blocks exist, or for rescue workers to know which houses
are on fire.

Because the notion of similarity is hard to specify precisely, even for manual
labeling, we  consulted  experts who have experience in disaster events. We
worked with members of the City of Berkeley's Disaster and Fire Commission, and
the local Amateur Ham Radio Club \cite{fire-commission}, who help local
neighborhoods during fire and earthquake events. They labeled the data
themselves, and also trained two of us on how to do so.  We treat two images as
similar if the average score of all four labelers is at least $3$.


Although images are labeled 
\emph{pairwise}, when 3 redundant images are found to be all pairwise similar, only one of them needs to be transmitted.  
However,  that the notion of similarity is not necessarily transitive.  That is, the
labeler marked images $A,B$ and $B,C$ were similar but $A,C$ were not. Given these
considerations, we define the notion of a
\emph{maximally similar set} for quantifing the redundancy in a
dataset. 

Let $S$ be the set of all images.  Formally,  $S_i \subseteq
2^S$ is maximally similar if $\forall p_j, p_k \in S_i,
\mathit{similar}(p_j,p_k)$ and $\forall p_l \notin S_i, \exists p_i \in S_i,
\mathit{notsimilar}(p_i,p_l)$. Intuitively, sending one image $p \in S_i$
should suffice to represent the perceptual content in this set.  Note that this
definition  handles both concerns raised above;  sets of more than two images
with pairwise similarity and cases where the similarity is not transitive.

Then, we quantify  the amount of redundancy using the notion of a
\emph{minimum  set cover} over the maximally similar sets. That is, $C \subset
2^S$ \emph{covers} the information contained in  $S$ if  $\forall S_i, \exists
p, p \in C \wedge  p \in S_i$.  If  $C^*$ is  the minimum set cover, we
quantify the redundancy in $S$ as $1-\frac{|C^*|}{|S|}$.  To see why this is an
intuitively reasonable notion, consider the  two extreme examples:  if all the
images were duplicates of the same source image the redundancy would be
$\frac{N-1}{N}$ and if all images were unique then the redundancy would be 0.
Using this definition and our labeled data,  {\em SDfire} dataset has 53\%
redundancy, and the {\em Haiti} dataset has 22\% redundancy. This means that
sending less than half the photos in the {\em SDfire} data, we can convey all
of the unique information.



\subsection{Do Byte-Level Methods Suffice?}
\label{sec:comparison}

A natural question is whether existing byte-level compression methods can
capture the redundancy in these datasets. We consider two  methods --
gzip-based compression and chunk-level compression~\cite{lbfs,RE-sigcomm08}.
For gzip, since each image is already in a compressed encoding, we consider
compressing the entire set of images (i.e., tar + gzip).  For the chunk-based
compression, we vary the chunk size to explore the tradeoff between chunk size
and  redundancy.

\begin{table}[htbf]
\small
\begin{center}
\begin{tabular}{c|c|c}
Method & Haiti & SDfire \\\hline
tar+gzip & 7.7\% &  5\% \\
Chunk-based, 64B &  2.2\% & 0.9\% \\
Chunk-based, 512B &  0.67\% & 0.04\% \\
Chunk-based, 2KB &  0.6\% & 0\% \\
Ideal/Content-Aware & 22\% & 53\% \\
\end{tabular}
\tightcaption{Redundancy elimination gained using different
methods on the Haiti and SDfire datasets.}
\label{tab:redundancy}
\end{center}
\end{table}

Table~\ref{tab:redundancy} shows that existing byte-level approaches cannot
capture  the redundancy in the two image datasets from disaster scenarios.  The
main reason that these byte-level approaches fail is that images that are
visually similar have many small photometric differences, causing their
byte-level encoding to be very different.

This result shows that while this is a non-trivial opportunity to leverage
redundancy elimination for reducing network load in disaster regions, directly
applying existing techniques does not work. We note that both datasets capture
a lower bound of redundancy, as these are  taken by users under stress. Photos
taken by people in real stress are likely to exhibit higher redundancy, hence
enable better compression ratios than evaluated in this section.


\section{CARE Architecture}
\label{sec:arch}

This section highlights two key design choices of CARE. First, we observe that defining a network
architecture for disaster response presents a set of unique challenges. It is
hard to estimate both the degree of damage to the infrastructure and the
network demands given that all traffic and usage models derived during normal
operating conditions are not valid or applicable.  Consequently, we envision
the use of techniques from  delay tolerant networking (DTN)~\cite{dtn}. Second,
our previous measurements indicate opportunities for reducing the network load
using content-aware redundancy elimination. Thus, we describe how traditional
DTN protocols can be augmented with content-aware strategies.

\subsection{Setting}
The typical response in disaster-affected regions is to deploy mobile
infrastructure that temporarily patches the damaged infrastructure. An example
of such infrastructure is based on COWs (Cell-on-wheels) that are transported
with trucks (or even air-lifted in extreme cases) and provide cellular phone
coverage over the disaster area~\cite{fcc-report}.  Given the limited
bandwidth, these systems are usually restricted to first responders and do not
allow access to the general population. In contrast, we are interested in
supporting the communication needs of regular citizens. Our vision is that
citizens impacted by the disaster will be able to use the same smartphones or
laptops they use during everyday life.

\begin{figure}[h] \begin{center}
\includegraphics[width=0.45\textwidth]{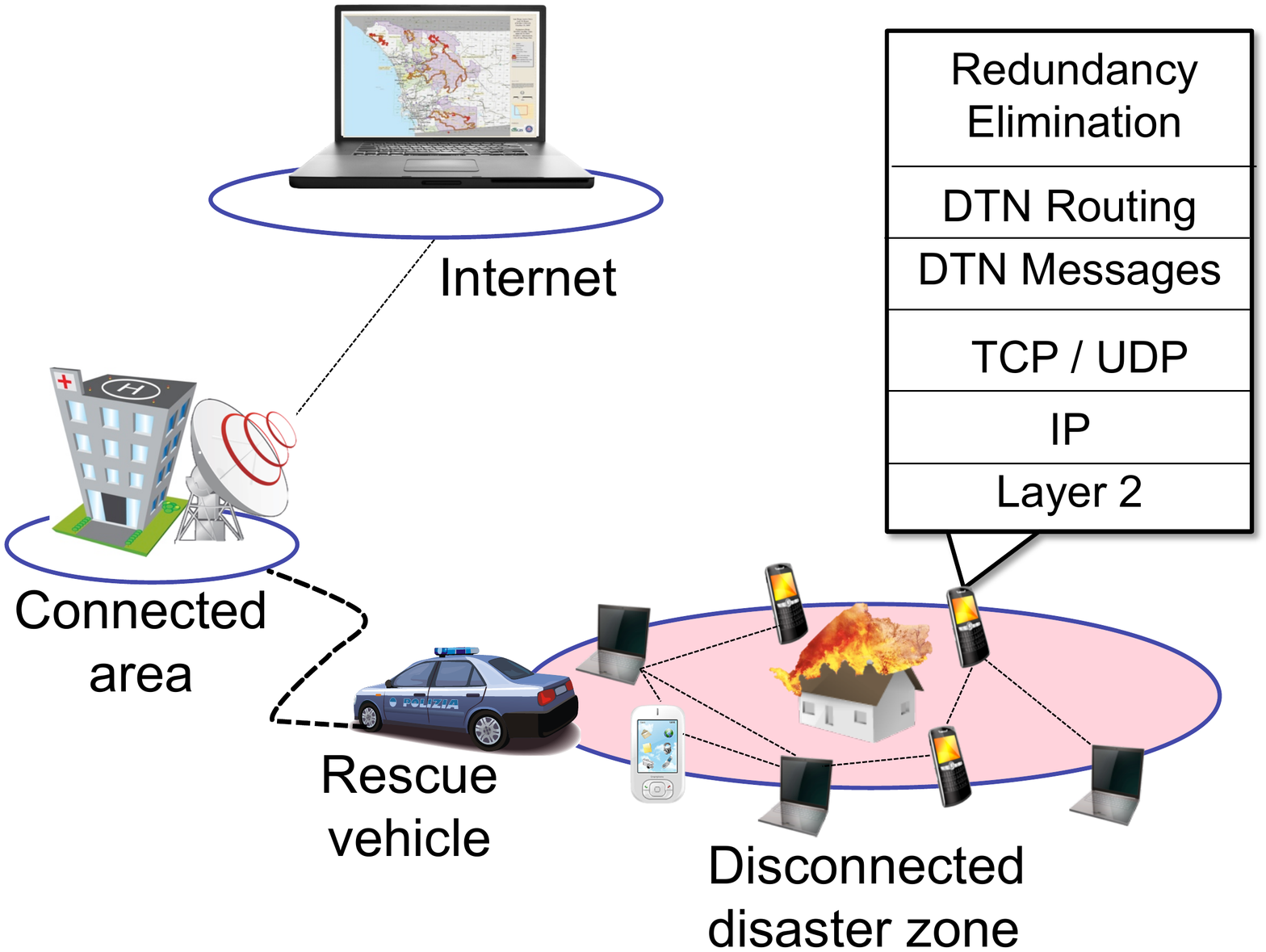} 
\end{center}
\vspace{-0.4cm}
\tightcaption{Example of a typical disaster scenario. Nodes in the disaster
region may have intermittent opportunities for external connectivity (e.g., a rescue vehicle).} 
\label{fig:scenario}
\end{figure}

Consider the scenario in \figref{fig:scenario}, where
there is no network
infrastructure available to users inside the area impacted by the disaster.
Users can reach any destination on the Internet as long as they manage to get
their messages out of the disaster area, via a device that can get in and out of the disaster area, which
in the figure is the rescue vehicle. Inside the impacted area the only
possible communication between users' devices is via ad-hoc or pairwise contact
opportunities.

We assume that a variety of devices (cellular phones, smartphones, laptops)
will be present in the disaster area. The devices will have a heterogeneous set
of network capabilities (WiFi, bluetooth), compute power (ranging from low
power embedded devices to multi-core processors), storage capacity
(from the 100s of Mbytes of a regular phone to the 100s of GBytes of a laptop)
and battery life (from a few hours to days). In practice, this means that nodes
can establish point-to-point communications, store large amounts of information
and are capable of processing the data they receive.

\subsection{Use of DTN}
In order to reach a destination outside of the disaster area (the ``outside
world'') the messages can only hop from one device to the other until one of
the devices is in range of the undamaged network infrastructure. This
communication model lands itself very well to the \emph{store-carry-forward}
 paradigm of Delay Tolerant Networks (DTNs)~\cite{dtn}. Hence, we propose a
software architecture in which devices, in absence of network infrastructure,
can enable a DTN stack -- a sort of \emph{disaster mode} for phones and laptops.
A DTN stack gives us mechanisms to discover neighboring nodes, identify the
available communication media, and to package, store and carry messages of
others.

The DTN stack, however, leaves open the choice of routing protocol.  There are
many proposals for DTN routing
protocols~\cite{jain.sigcomm04,lindgren.mobihoc03,epidemic,rapid}.  For our
specific environment we choose Epidemic Routing~\cite{epidemic}. In epidemic
routing, when two nodes come in contact, they  exchange all messages that
they do not already possess.  
 We choose epidemic routing for three main reasons.
First, message latency is a critical performance metric in disaster scenarios.
Depending on the content of the message it may really be a matter of life and
death. As long as there are enough buffers to store incoming messages, epidemic
routing guarantees optimal latency. Second, we cannot make any a-priori
assumptions on the mobility of users or on the presence of regular patterns
(e.g., bus schedules).  Different types of disasters impose very different
restrictions on the movement of people and cause different level of damage to
the nodes.  Thus, it rules out several DTN routing protocols that leverage such
assumptions for more efficient message delivery.  Third, many DTN routing
proposals are tuned toward reaching a specially marked destination node.
However, within a disaster area, all messages are really destined to the
``outside world'' rather than a specific destination.  Once outside the
disaster area the regular Internet protocols will carry the messages to the
intended destination. In this case, epidemic routing maximizes the chances to
reach the unknown and possibly variable set of nodes with connectivity to the
regular infrastructure.

At the same time, we note that the content-aware strategies that we propose
next are not tightly coupled to the epidemic routing. In other words,  they can
be used in conjunction with, and provide benefits for, other routing
strategies.


\comment
{
One of the side effects of epidemic routing is that it increases the number
of messages that are transmitted in the network. In a situation where anxious
users are prone to attempt to communicate more than usual, this will only
exacerbate rather than alleviate the problem of operating with a damaged
network infrastructure.
}
\subsection{Augmenting DTN with Content-Awareness}

In order to limit message replication and reduce 
network load, we use the message content to drive our forwarding decision. Our
system, called CARE (Content Aware Redundancy Elimination), detects when a message
is semantically similar to another message, hence de-prioritize one of them
and transmits the other. De-prioritizing may mean pushing that
message (or ``bundle" in DTN parlance) to the back of the transmission queue or
even marking it as ``first-to-discard'' if memory buffers are full. To simplify
the discussion, we assume that redundant content is dropped. 

A different form of prioritizing messages was proposed in MaxProp DTN routing 
protocol \cite{maxprop}, and was shown to out perform other DTN protocols. However,
MaxProp is based on the path likelihoods of nodes to a known destination based on historical
data. The assumption of predictable paths is not applicable in our disaster area setting.

The overall goal of our system is to ensure that maximum amount of unique
information is transmitted from the disaster zone. In this regard, the novel
aspect of our design is to to incorporate content-aware redundancy elimination
into the DTN routing protocol. 

\ignore{
Without any mechanism to identify redundant content, overflowing buffers simply
drop potentially important messages in a FIFO manner. The overall goal of our
system is to ensure that maximum amount of unique information is transmitted
from the disaster zone. One of the novel aspects of our design is the proposal
to incorporate redundancy elimination at the core of the DTN routing
protocol.  When two nodes enter contact range they exchange only ``unique''
messages in their transmit queues.
}

\begin{figure}[h]
\begin{center}
\vspace{-0.5cm}
\includegraphics[width=0.45\textwidth]{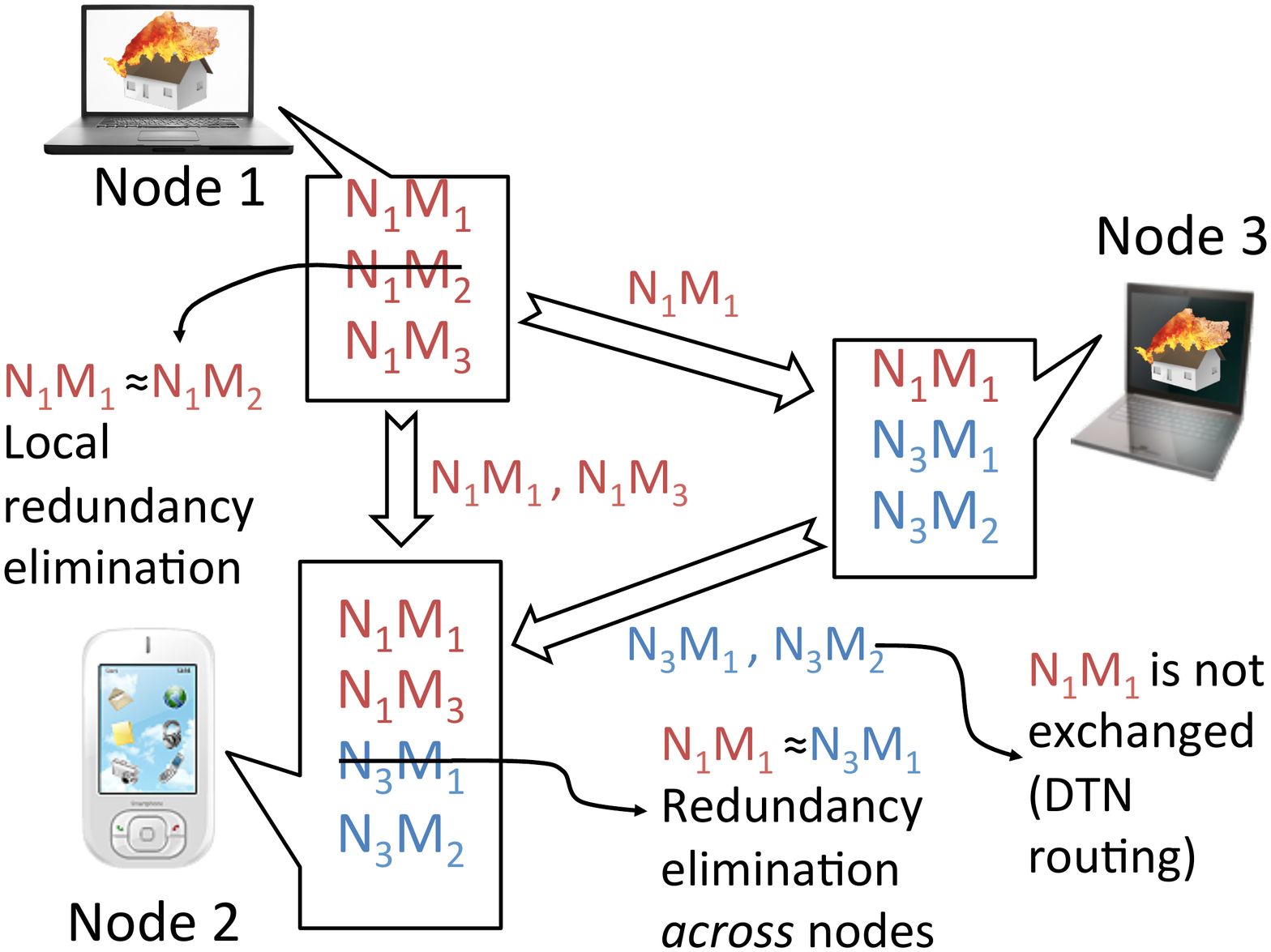}
\end{center}
\vspace{-0.5cm}
\tightcaption{Augmenting DTN routing with content-awareness.}
\label{fig:exchange}
\end{figure}


Traditional DTN routing uses globally unique message ids to identify repeated
content and avoids re-sending them. As \figref{fig:exchange} shows, CARE augments
DTN routing in two ways.  First, it enables a node to detect local
redundancy. Node 1 detects that message $N_1M_2$ is
identical to $N_1M_1$ and discards it. Node 1 then contacts Node 2
and sends only messages $N_1M_1$ and $N_1M_3$. It then comes in range with
Node 3, but manages to deliver only $N_1M_1$, e.g., due to a short contact time. 

Node 3 generates its own messages $N_3M_1$ and $N_3M_2$.
When it contacts Node 1, it only sends $N_3M_1$ and $N_3M_2$, since DTN routing identifies
that $N_1M_1$ already exists by comparing id. CARE further enables Node 2 to detect that message
$N_1M_1$ is similar to the newly received $N_3M_1$, therefore it drops the latter, saving buffer space
and reduces message flooding.

Overall, the proposed CARE architecture has the following advantages over standard DTN routing:
\begin{packeditemize}
\item Redundant messages are first identified locally, thus avoiding further propagation of redundant content.
\item CARE extends id-based duplicate detection with semantically redundant message detection across different nodes.
\item Without any mechanism to identify redundant content, overflowing buffers
drop  messages in a FIFO manner. In CARE, redundant messages are dropped first
enabling storage of unique messages for longer durations.  Thus, more unique
messages reach the destination.
\item The power saved by lowering the number of transmitted
messages typically exceeds the power consumed for detecting
redundancy~\cite{energy}. Prolonging the battery life of the device by enabling
efficient communication can have direct impact on the person using it during
the disaster.
\end{packeditemize}

\ignore{
An additional benefit which we gain by dropping redundant
content, is the reduction of the number of messages that needs to be
transmitted between devices. The large number of messages generated during the
disaster  has a severe impact on the power consumption of the devices. Although
our proposed algorithms consume CPU cycles, the power savings made by not
sending a message exceeds by far the power consumed by the CPU~\cite{energy}.
Prolonging the battery life of the device by enabling efficient communication
can have direct impact on the person using it during the disaster.
}



\section{Content-Aware Image Traffic Reduction}
\label{sec:algo}

This section provides a brief description of the methods we use to detect that
two images are ``similar enough" so that one of them can be dropped. We
describe and evaluate 3 computer vision techniques. Subsequently we design a
hybrid method that tries to leverage the advantages of each while minimizing
their drawbacks.

\ignore{
In general, the process of detecting redundant images contains two main steps. The first step is to find a compact way to represent each image. One approach (often called ``global method") is to analyze features of the entire image without segmenting it. Example methods of this type include perceptual hashing (pHash) \cite{robust-phash} and GIST for scene detection \cite{gist}. Alternatively, one can partition the image into smaller portions and extract local features from each portion. A popular method of this type is SIFT \cite{sift}. The second step is to use the image representations extracted in step one and to compare them. The comparison assigns a distance measure between  the two images.This comparison is highly dependent upon the method used in step one.
The global methods used in step one are typically faster yet less accurate than the local methods which incur the opposite tradeoff. Methods like SIFT are also more robust to photometric and geometric variations than global methods.
}

\subsection{SIFT}

Scale-invariant feature transform (SIFT) \cite{sift} is an algorithm that finds
a representation of an image that captures consistent properties that are
invariant to image translation, scaling, and rotation, and some illumination
changes \cite{csift,vandeSandeTPAMI2010}. The algorithm first  finds
interesting regions or {\em keypoints}, each of which is represented by a
feature vector. A keypoint is a region that might be identified because it has
a clear edge boundary, or because it is typically either brighter or darker
than its neighborhood. A typical image is represented using several thousands
of feature vectors.  Given this representation, SIFT uses feature matching for
image similarity detection. For each keypoint in the first image, it finds a
matching ``nearest'' keypoint in the second image based on the Euclidean
distance between their feature vectors.

Consider two images $i$ and $j$ with corresponding feature vectors $v_i$ and $v_j$. SIFT finds $m$ keypoints that
match (seem to capture the same region), and a smaller set  $m' \leq m$ of keypoints that are most likely to be
the same only viewed from different angels. The SIFT similarity score $S_{sift}$ is then calculated using:
$$S_{sift}=\frac{m'}{m}$$

\noindent If  $m>8$ (a common SIFT configuration) and $S_{sift} > T_{sift}$ then two images are identified as redundant.

\subsection{pHash}


A perceptual hash (pHash) \cite{robust-phash} is a fingerprint of a multimedia
file derived from the content.  pHash produces one hash for the entire image
without decomposing the image into smaller elements, making it computationally
cheap to run. pHash is commonly used for copyright violation detection because
it is good at detecting when two images are nearly identical, despite the
introduction of small variations such as watermarks and minor transformations.
In particular, it generates a $R=64$-bit feature vector using the  Discrete
Cosine Transform (DCT) of the image.  Let $p_i$ denote the pHash vector for an
image $i$. The similarity between two images is calculated as $$S_{ph} = R -
HammingDist(p_1,p_2)$$

\noindent where {\em HammingDist} is the usual Hamming distance, i.e., the
number of similar corresponding bits between the hashes.  We say that two
images are redundant if $S_{ph} > T_{ph}$ where $T_{ph}$ is the cutoff
threshold for pHash.

\subsection{GIST}
GIST \cite{gist} is a method used for scene detection. A scene is defined as an
image that has roughly at least 5 meters between the observer and the focal
point. This differs from images of ``objects" in which the objects are often at
a hand's distance from the observer. GIST provides a representation of a scene
by using spectral and coarsely localized information to describe the ``shape of
the scene". GIST uses a set of perceptual dimensions (naturalness, openness,
roughness, expansion, ruggedness) to capture the dominant spatial structure of
a scene.

We include GIST in our study because it has the potential to be useful in
disasters involving fires, floods, or tornados. Photos from such events could
be of scenes with amorphous elements (such as a spreading fire) that don't have
clearly well defined edges (something that is important for methods such as
SIFT).

Let $v_i$ and $v_j$ be the GIST vectors representing images $i$ and $j$,
respectively. The similarity measure of the two images is calculated as:
$$S_{gist}=\frac{cov(v_i,v_j)}{\sigma(v_i)\cdot \sigma(v_j)}$$

\noindent where $cov$ is the covariance function and $\sigma$ is the standard deviation function. We consider two images to be similar if $S_{gist} > T_{gist}$, where $T_{gist}$ is the cutoff threshold for GIST.

\subsection{Performance of Individual Methods}

We now analyze the performance of these image similarity detection methods on our datasets. The configuration of each method, i.e. the setting of the threshold parameters $T_{sift}$, $T_{ph}$, $T_{gist}$ will determine its false positive and false negative performance. When the detector makes a false negative (FN) mistake, it means that it failed to detect that two images are indeed similar. A false positive (FP) mistake means that two different images were considered similar. From the perspective of our SA application, an FP mistake is far worse than an FN one, because an FP mistake means that a piece of unique information will get dropped (or put in the back of the queue) and might never reach the SA service. An FN mistake merely means that our use of the available bandwidth is less efficient than optimal because the network will carry a piece of redundant content.

To understand the FP and FN tradeoffs for these 3 methods we generated ROC curves for each of them. Figures ~\ref{fig:roc_haiti} and ~\ref{fig:roc_sdfire} show the resulting curves when applied to the {\em Haiti} and {\em SDfire} datasets respectively. Each point on the SIFT curve corresponds to a particular threshold, $T_{sift}$, and we sweep through many values to generate the full curve. The pHash and GIST curves are created similarly by varying $T_{ph}$ and $T_{gist}$, respectively. For each threshold value, we examined all pairs of images to determine whether or not the algorithm deems them similar. Accuracy is computed against our ground truth labels (from Sec. \ref{sec:image_redundancy}).

\begin{figure}[tbh]
\centering
    \subfloat[{\em Haiti}]{
	\label{fig:roc_haiti}
	\includegraphics[width=0.40\textwidth]{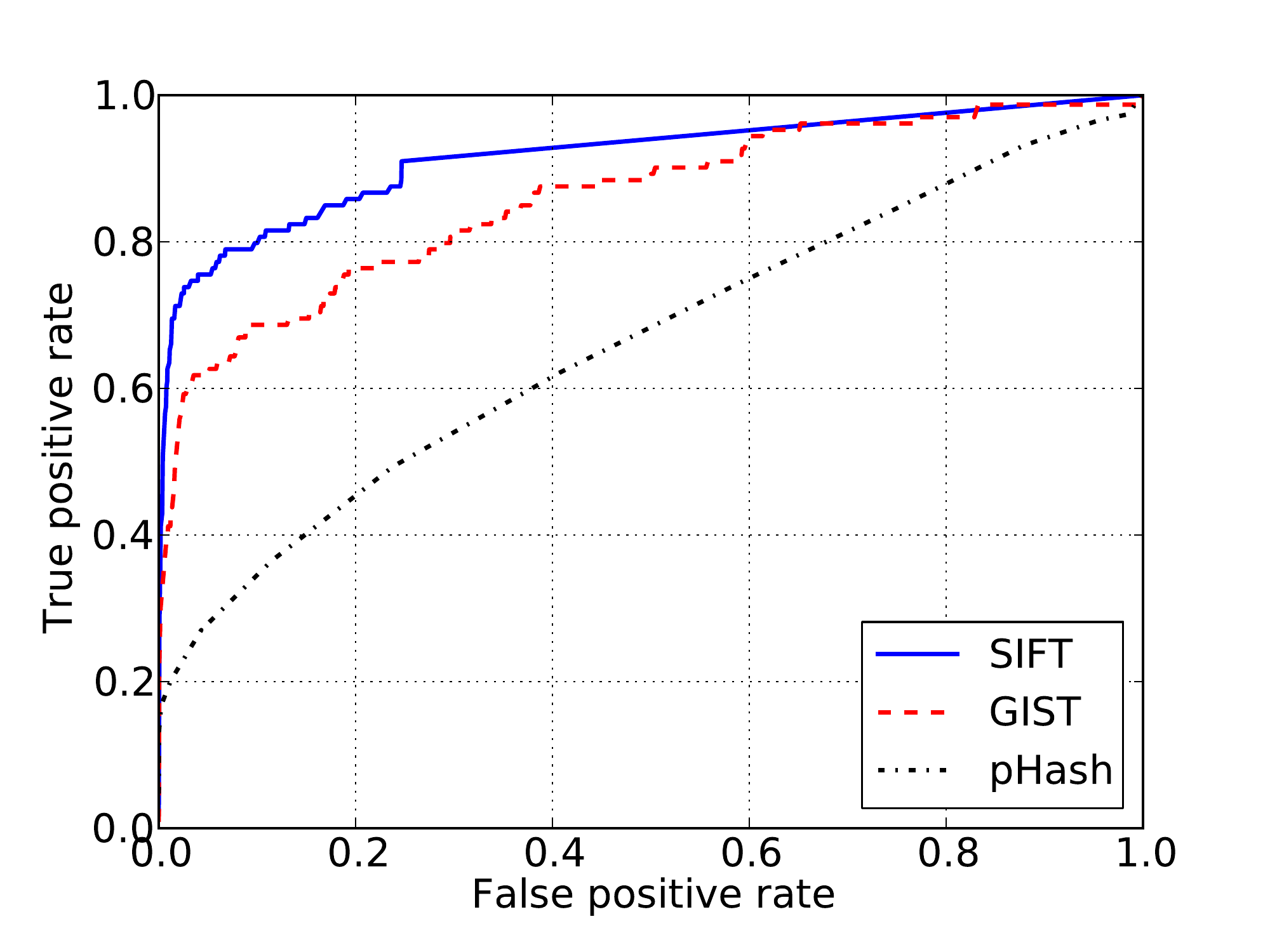}
	\vspace{-0.3cm}
    }
    \hspace{-2mm}
    \subfloat[{\em SDfire}]{
    \label{fig:roc_sdfire}
	\includegraphics[width=0.40\textwidth]{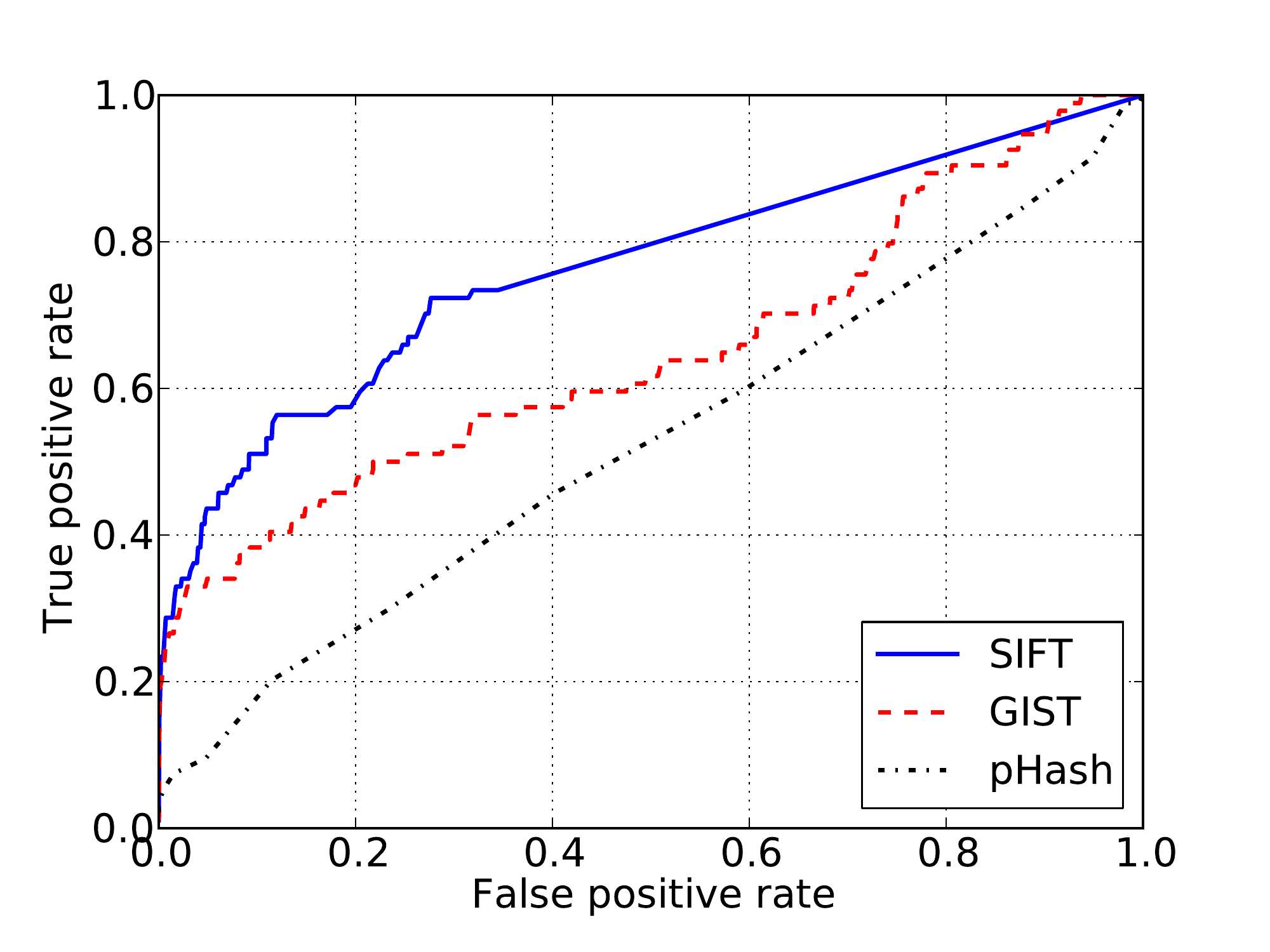}
    	\vspace{-0.3cm}
    }
    \tightcaption{ROC curves for three image similarity identification methods}
    \label{fig:rocs}
\end{figure}

\ignore{
\begin{figure*}[tb]
\begin{center}
\begin{minipage}[t]{.45\textwidth}
\centering
\includegraphics[width=\linewidth]{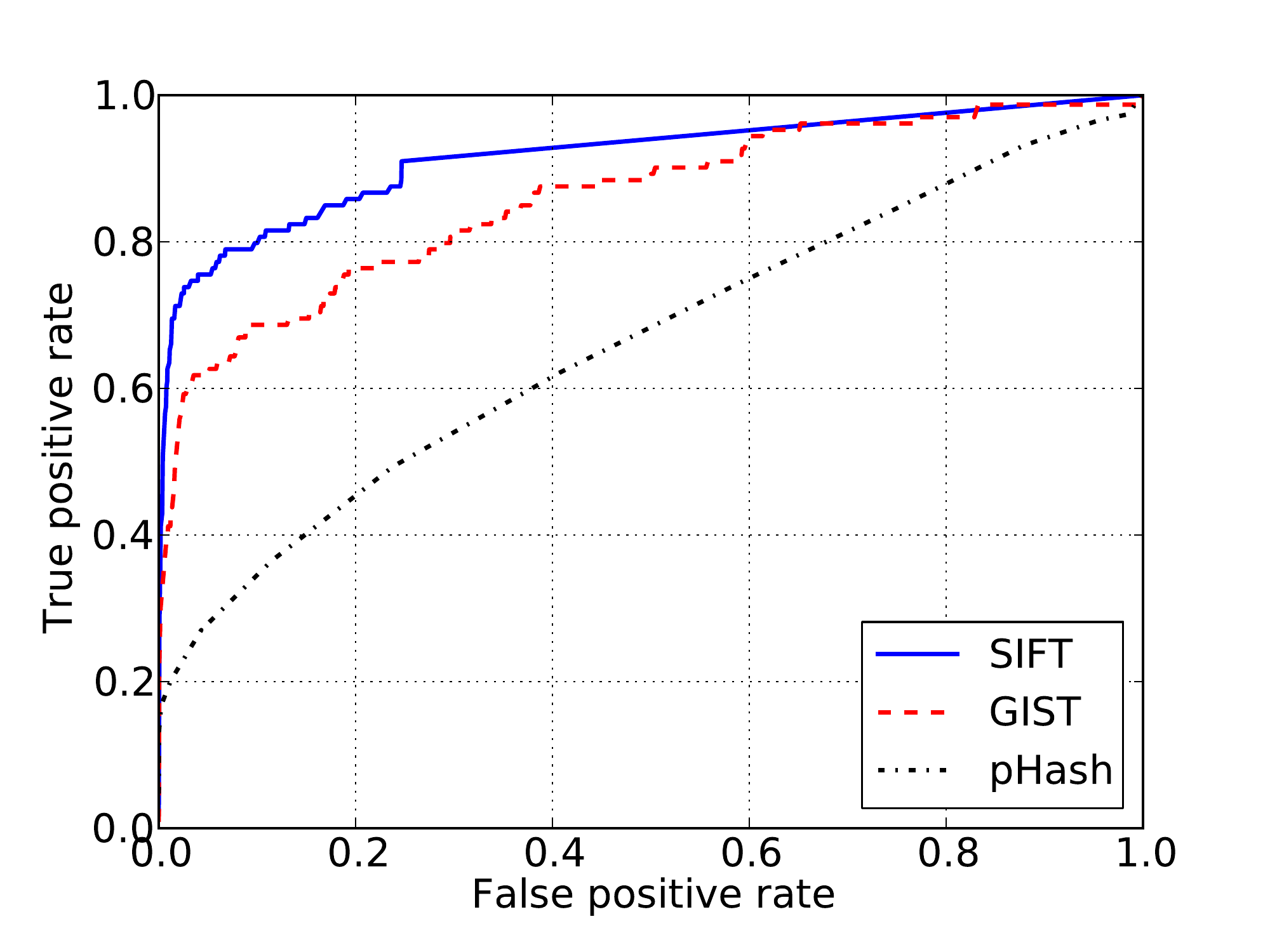}
\tightcaption{ROC curves for 3 image similarity identification methods on Haiti dataset}
\label{fig:roc_haiti}
\end{minipage}\hspace{1.0em}
\begin{minipage}[t]{.45\textwidth}
\centering
\includegraphics[width=\linewidth]{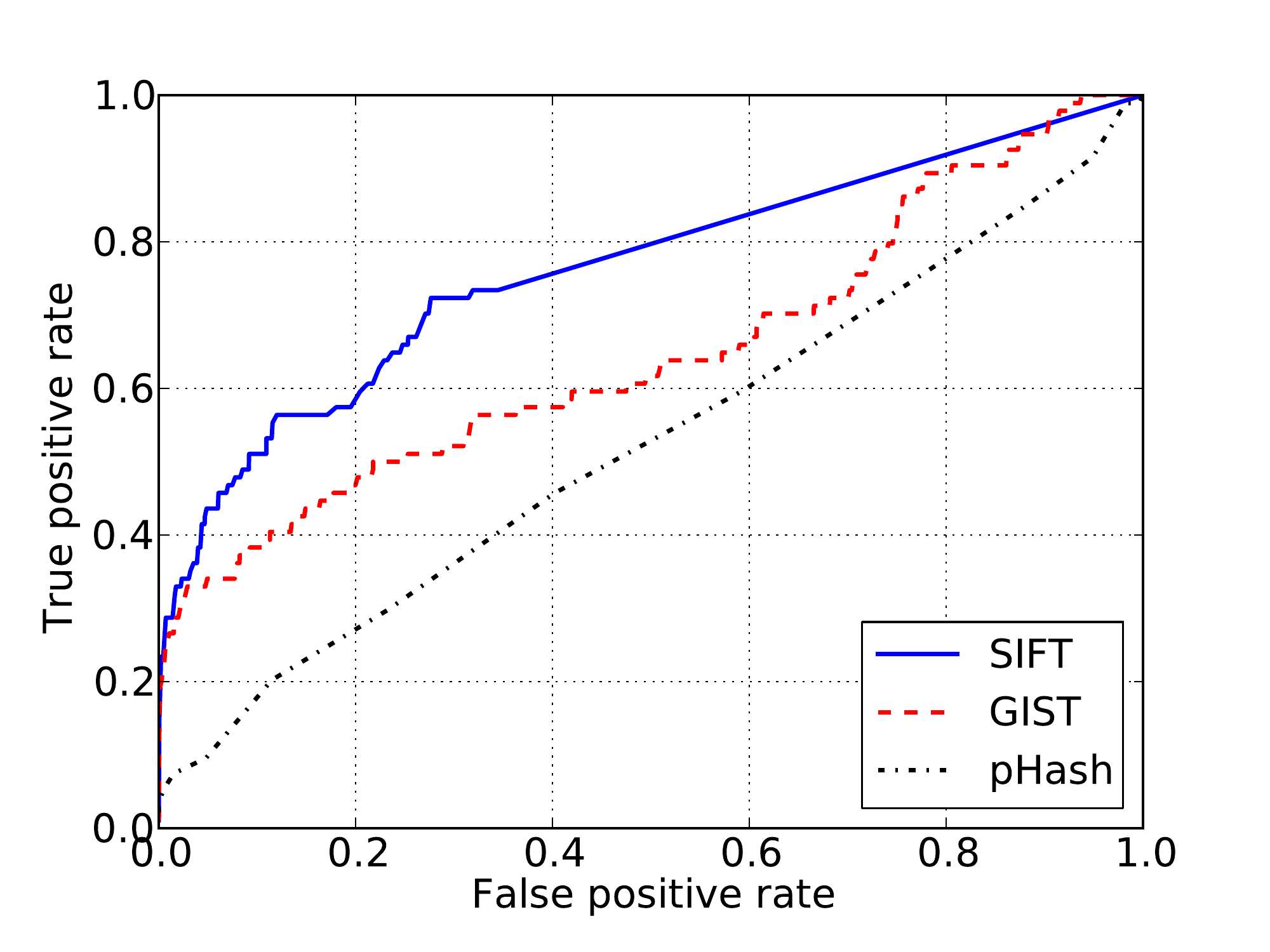}
\tightcaption{ROC curves for three image similarity identification methods on SDfire dataset}
\label{fig:roc_sdfire}
\end{minipage}
\end{center}
\vspace{1em}
\end{figure*}
}

\figref{fig:roc_haiti} shows that for the Haiti dataset both SIFT and GIST performed reasonably well. With an FP rate of 5\%, SIFT can correctly identify over 75\% of all true matches and GIST can identify roughly 60\%. Although the performance of pHash is not superb, this is expected since pHash is only good at identifying similarity when two images are nearly identical. pHash is not robust to geometric transformations, distortions in luminance between images, nor images with only partial overlap. We note that for a low FP rate of 1\%, pHash can identify roughly 21\% of the redundant image pairs. In the next section, we see how this can be leveraged when we design our hybrid method.

\begin{figure*}[tbh]
\centering
    \subfloat[]{
	\label{fig:all-success}
	\includegraphics[width=0.30\textwidth]{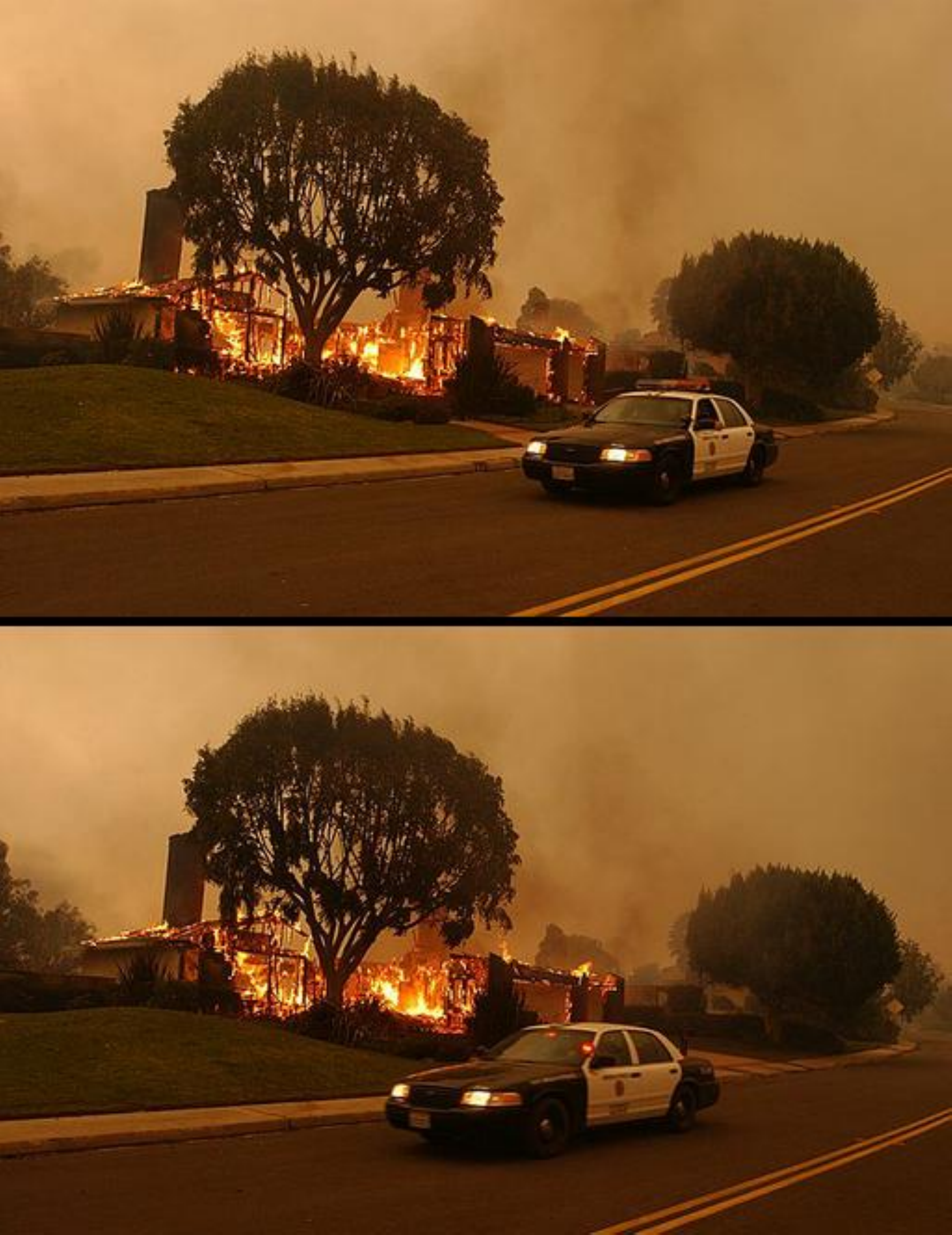}
    }
    \subfloat[]{
    \label{fig:gist-fails}
	\includegraphics[width=0.30\textwidth]{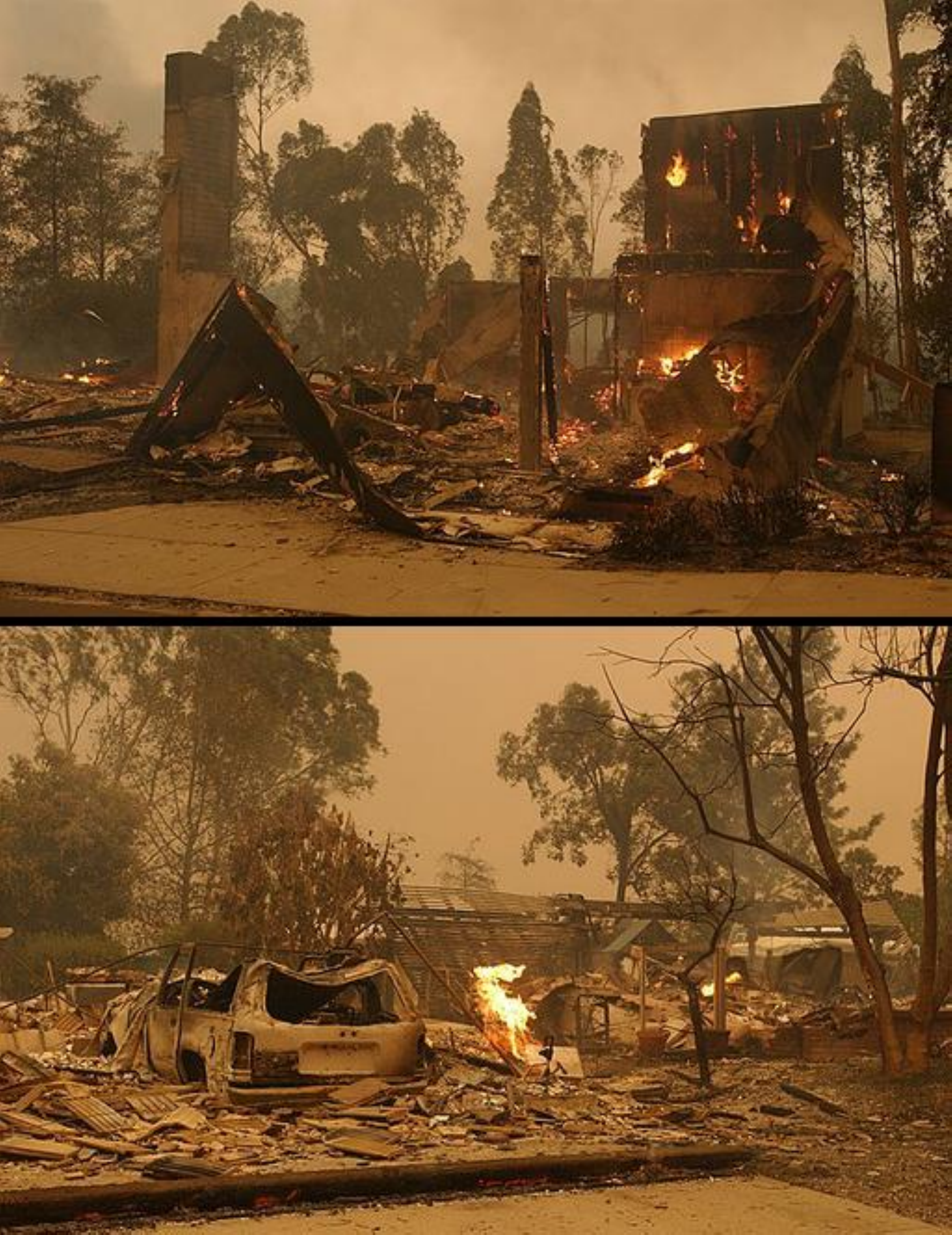}
    }
    \subfloat[]{
    \label{fig:sift-fails}
	\includegraphics[width=0.30\textwidth]{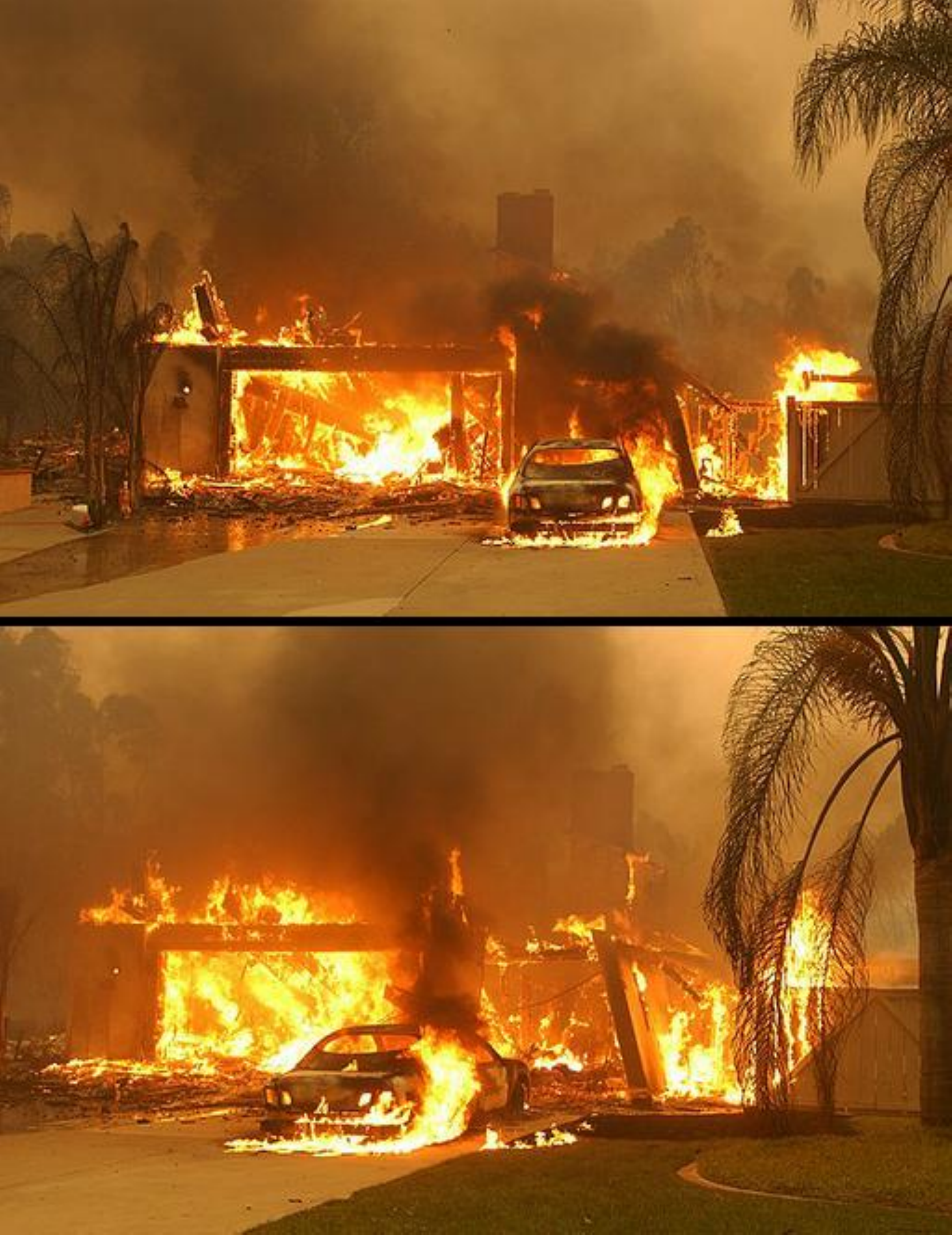}
    }
    \tightcaption{Examples of image pairs in the SDfire dataset, showing (a)  an image pair that is successfully identified as redundant by all methods, (b) an image pair that GIST incorrectly identifies as redundant, due to the apparent similarity of the scenes, and (c) an image pair for which SIFT fails to identify the redundancy, due to the moving boundaries of the fire and smoke}
    \label{fig:examples}
\end{figure*}

In \figref{fig:roc_sdfire} we see that these methods do not perform as well on the {\em SDfire} data. At a 5\% FP rate, SIFT cannot do much more than detect about 42\% of all the true matches, while GIST correctly identifies about 35\%. pHash performs poorly. We believe that SIFT performs better on the {\em Haiti} than the {\em SDfire} data because the {\em Haiti} photos often capture an urban area, or areas with well defined objects such as roads, cars, people, or buildings. The fires and smoke in the {\em SDfire} data result in scenes whose elements are without clear edges. We suspect that GIST has trouble here because the images are not ``pure" scenes but rather scenes often with a small number of well defined objects. \figref{fig:examples} provides an illustration of the kinds of image pairs for which these techniques either succeed or fail.

Overall,  SIFT outperforms the other  methods. However SIFT is also
computationally more expensive.  Since handhelds during a disaster may be
constrained by power, the cost of these algorithms is a
factor to consider when putting such methods in the network layer.

In order to better understand their costs, we measured the execution times
of the three algorithms on two very different platforms~\cite{ark}:
$(i)$ a $1.6$~GHz Atom
330 board with $2$~GBytes of RAM, and, $(ii)$ an 8-core $2.8$~GHz
Xeon-based high end server with $12$~GBytes of RAM. We ran the three
algorithms on pairs of images in our datasets using the same (unmodified) code
on the same operating system on both platforms.
Of course, the execution times on the two platforms are very
different but the absolute values are not that useful in guiding design
decisions in our system. In fact, the actual execution times are bound to
change drastically every year as
a new generation of processors and devices enters the market.
Instead, a metric that can help us in the design of our system is the
ratio of execution times and whether those ratios change significantly across
platforms. The result of our investigation is that,
as expected, SIFT is the most expensive algorithm to run: the execution time
of SIFT is on average 150x that of pHash while GIST is 50x that of pHash. We also found
that these ratios are quite consistent across the Atom and Xeon-based platforms.



\subsection{Pipeline Method}

\begin{figure}[h]
\begin{center}
\includegraphics[width=0.45\textwidth]{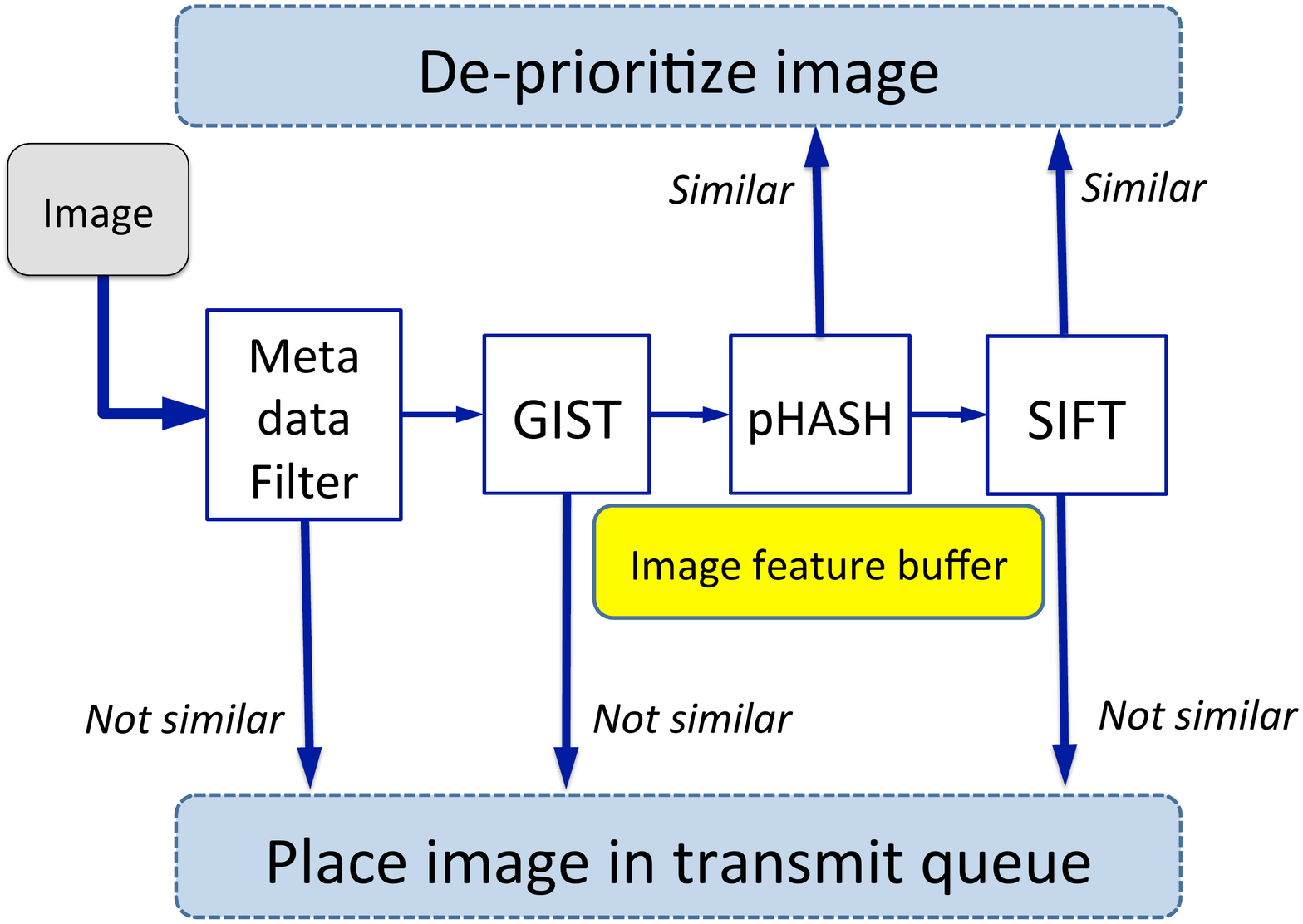}
\end{center}
\tightcaption{Pipeline Method}
\label{fig:pipeline}
\end{figure}

In order to balance the tradeoffs between FPs, FNs and cost, we designed a hybrid method that combines these algorithms in a pipeline as depicted in \figref{fig:pipeline}. We were motivated by cost issues, but also because these 3 algorithms have different strengths and weaknesses, and we believe we can leverage the strengths to reduce cost, while diminishing some of the weaknesses.

We point out some behaviors of GIST and pHash that affected our pipeline design.
GIST sometimes wrongly identifies non-similar images as being similar, especially when the
scenes are similar but the details within them are different (see \figref{fig:gist-fails}). However,
since non-similar images often capture different scenes, GIST actually does well at identifying when two images are {\em not similar}. In other words, it is easy to find a threshold $T_{gist}$ such that when $S_{gist} < T_{gist}$ there is a high probability that the images are not similar.

pHash can only be used to identify images that are obviously similar (such as exact duplicates, or images that only differ by a small amount of lighting or focus) and we can avoid the heavy machinery of SIFT on the easy cases. (Although pHash exhibits weak performance on our datasets, we note that these are only two datasets and much remains to be learned about SA datasets from disasters. We suspect that datasets of the future could easily contain a great deal of near-duplicate photos, and thus we include pHash in the general design of our pipeline method.)

Because of these properties, we designed our pipeline method as follows. A node is comprised
of a single limited buffer that holds feature vectors of images previously received by the node and an outgoing message queue. When a new image arrives, the pipeline first tests if it has metadata, such as time or
geo-tagging. Geo-tagging, using GPS or other forms of positioning
is becoming increasingly common in modern cameras and handhelds. For example, in~\cite{BuildingRome}
the authors report that roughly 10--15\% of their city photos obtained from Flickr are geo-tagged. The pipeline uses this data to decide whether the
image was
taken too far away or too long ago from any other image in its buffer, making an easy dissimilarity decision, thus
adding it to its send queue. Otherwise, the pipeline needs to run the algorithmic phases.

GIST first compares it to other images in the buffer. Because we only use GIST here to detect {\em non-similarity}, if the image is different from all those in the buffer, then it is added to the send queue. Otherwise, we assume GIST cannot precisely identify whether the images are similar, hence it forwards it to the next stage in the pipeline. By using GIST to detect non-similarity, it means that all other decisions (i.e., about similarity) are deferred to stages further down the pipeline. This means that the pipeline incurs any false negative mistakes GIST may make, but it does {\em not} incur the FP mistakes GIST would normally make. Since decisions about similarity are deferred down the pipeline, the impact on FPs of the overall system will be influenced only by pHash and SIFT (that do make decisions about similarity). We put GIST first in the pipeline because the majority of the pairs in our datasets are not redundant, and this reduces the overall cost since most images will not have to be processed by SIFT.

The next phase uses pHash only to detect similarity. pHash compares the current image to those in the buffer, and if it decides it is similar, then the image is dropped or de-prioritized. We configure pHash for a low FP rate, because its FN behavior does not impact the pipeline's overall performance since pHash makes no decisions about non-similarity. If a decision still has not been taken, SIFT processes the image and makes a final decision regarding the redundancy of the current image with those in the buffer. Intuitively, we are using GIST and pHash to handle the easy cases (of non-similarity and similarity, respectively), and we only use SIFT for the non-obvious cases.

We also considered another hybrid method based upon using a decision tree. Using the results of each method individually, and the ground truth labels, we ran the WEKA data mining tool \cite{weka} to generate an optimal decision tree. This method yielded an insignificant improvement in the FPs and FNs. Because SIFT is the most accurate method, it was elected to be at the top of the decision tree, which means that it needs to run on all pairs mitigating possible improvements in cost. We therefore decided not to pursue this method.

Our goal is design a pipeline that achieves similar or better accuracy (i.e.,
FP,FN) at a lower cost, than the best single method.

To study the performance of the pipeline we considered some sample scenarios on our Haiti dataset. Recall that in our pipeline method, we select thresholds for GIST based upon a target FN rate (since it only decides about non-similarity) and for pHash based on a target FP rate (since it only decides on similarity). SIFT is configured for a target FP rate because we consider FPs more important than FNs in the overall system. Hence we found thresholds so that $FN_{gist}$=30\%, $FP_{pHash}$=1\% and $FP_{sift}$=1\%.

Table~\ref{tab:pipeline} shows the FP, FN and cost of each of the individual methods and of the pipeline method. We normalize the cost by the cost of running pHash.  Let $N_i$ be the number of image comparisons conducted at stage $i$. Let $c_i$ be the cost of running the algorithm at stage $i$. Then the average cost of similarity identification using the pipeline is given by $(N_1 c_1 + N_2 c_2 + N_3 c_3) / N_1$ (note that $N_1$ is the total number of image pairs that need to processed in any singleton method). In the upper part of the table, we see that the pipeline method's FP and FN performance are close to SIFT (the best singleton method), yet the pipeline incurs about half the cost. We also consider a slightly different scenario in which pHash and SIFT are configured as before, but GIST is now configured for a $FN_{gist}$=10\% (GIST-2 and Pipeline-2 in the table). In this case the pipeline exceeds the performance of any single method, however the cost gain is more modest (roughly 15\% improvement).

\begin{table}[t]
\small
\begin{center}
\begin{tabular}{c|c|c|c}
Method &  FP & FN & Normalized cost \\
\hline
pHash &  0.01 & 0.83 & 1 \\
GIST  &  0.15 & 0.30 & 50 \\
SIFT &  0.01 & 0.48 & 150 \\
Pipeline  &  0.03 & 0.46 & 70 \\ \hline
GIST-2 &  0.55 & 0.10 & 50 \\
Pipeline-2 & 0.01 & 0.37 & 130 \\
\end{tabular}
\tightcaption{Comparison between the pipeline approach and the individual
algorithms.}
\label{tab:pipeline}
\end{center}
\vspace{-0.2cm}
\end{table}




\section{Disaster Scenario Simulation}
\label{sec:eval}


In order to understand the performance of CARE in a disaster scenario, we used the Opportunistic Network Environment (ONE) \cite{theone} simulator, which is a DTN simulator. We augmented this tool with an implementation of CARE, and a message generation process that allows us to control the amount of redundancy in the data -- thereby enabling sensitivity analysis.

\subsection{Scenario Settings}

We chose to simulate a disaster scenario in Pittsburgh,  USA because the ONE
simulator comes with a detailed map of Pittsburgh,\footnote{The Pittsburgh map
was contributed to The ONE by PJ Dillon,
\url{http://www.cs.pitt.edu/~pdillon/one/}} that includes all the roads and
bus routes. We focus  on a neighbourhood called Oakland, that covers
roughly an area of 10 miles by 8 miles. We consider a scenario with 50 people
randomly located inside the disaster area. A single rescue vehicle travels
between the disaster area and a communication gateway (in a different part of
the city), that represents the only device that has Internet connectivity
(e.g., satellite), and is located roughly 6 miles outside of the disaster area.
Although we term the vehicle traveling in and out of the disaster area a {\em
rescue vehicle} we note that it could simply be an individual in the area who
has the means and potential to move in and out of the disaster zone.

We generate messages ourselves in the simulator, rather than feed in our
datasets as input, because we wish to control the level of redundancy for
sensitivity evaluation. At each epoch, the simulator randomly selects a person
that generates a 300KB message with a unique identifier, targeted towards the
communication gateway. 300 KB represents a typical photo at a relatively low
resolution. A message is generated once every $G$ seconds, so that during a
simulation of $T$ seconds there will be $T/G$ messages generated. Let $M_i$
denote a message $i$, and $R_{sim}$ denote the percent of message redundancy in
the set of overall messages produced during an experiment. When the simulator
decides to add a redundant message it selects a random message $M_i$ at a
random user, and places the redundant message in the user's message stream as
though it occurred shortly before or after $M_i$ was generated. In particular,
we use a time window that started 20 seconds before $M_i$, or that ended 20
seconds after $M_i$. This allows us to simulate  temporal
locality, which we expect will occur often in such disaster scenarios.

People walk
at a speed uniformly distributed in the range of 3--7 Km/h (2--5 mph). The people move in a point-of-interest
map-based shortest path traversal, meaning each person selects a destination point inside the disaster area and finds
the shortest path using roads to get there. Once the person reached its destination it stops for 5 minutes, and then
repeats this process.

Each person has a WiFi-enabled device, transmitting to a maximal range of 20 meters in 10Mbps, simulating
a smartphone. This device also has a limited buffer used for storing messages
that the person creates or receives from others. We denote the size of the buffer in the people device by $B_{people}$.

The rescue vehicle drives at a speed uniformly distributed in the range of 25--54 Km/h (15--33 mph). The rescue vehicle selects whether to go to the disaster area or go back to the communication gateway. We denote the probability of selecting a destination within the disaster area as $Pr_{disaster}$, which is independent from the
current location of the rescue vehicle. Once selected to go to the disaster
area, the rescue vehicle selects a random point within it, drives there, and
once reached, it waits 5--10 minutes at that location before
selecting a new destination.

The rescue vehicle is equipped with a 100Mbps WiFi device, also transmitting to a maximal range of 20 meters. This device
enables the rescue vehicle to communicate messages with people at a maximal bandwidth of 10Mbps and with the
communication gateway at a maximal bandwidth of 100Mbps. We denote the size of the buffer of the rescue vehicle as $B_{rescue}$.

\begin{figure*}[tbh]
\centering
	\subfloat[Unique messages]{
	\label{fig:sim-run}
	\includegraphics[width=0.34\textwidth]{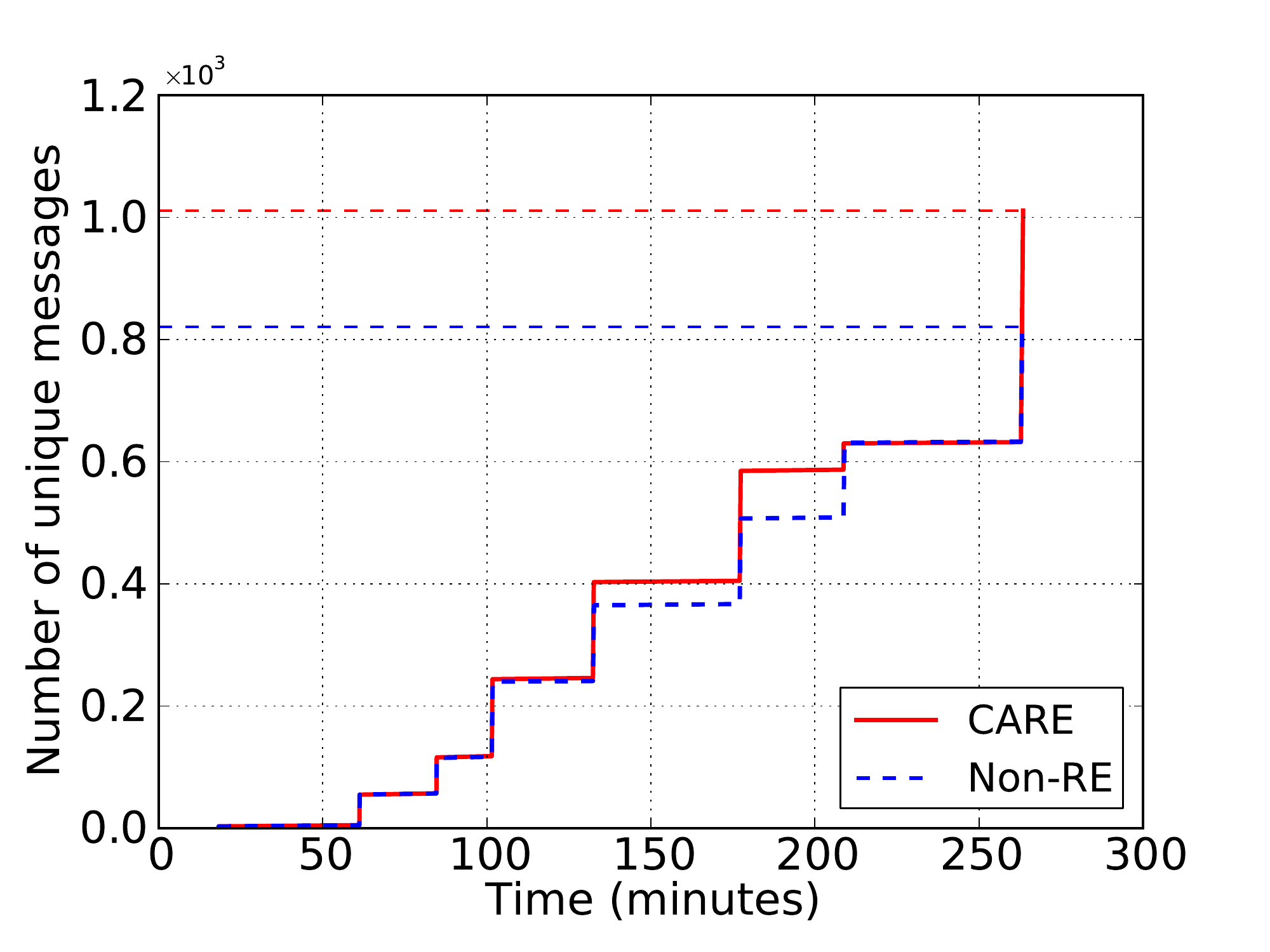}
    }
    \hspace{-7mm}
    \subfloat[Improvement]{
	\label{fig:imp-vs-time}
	\includegraphics[width=0.34\textwidth]{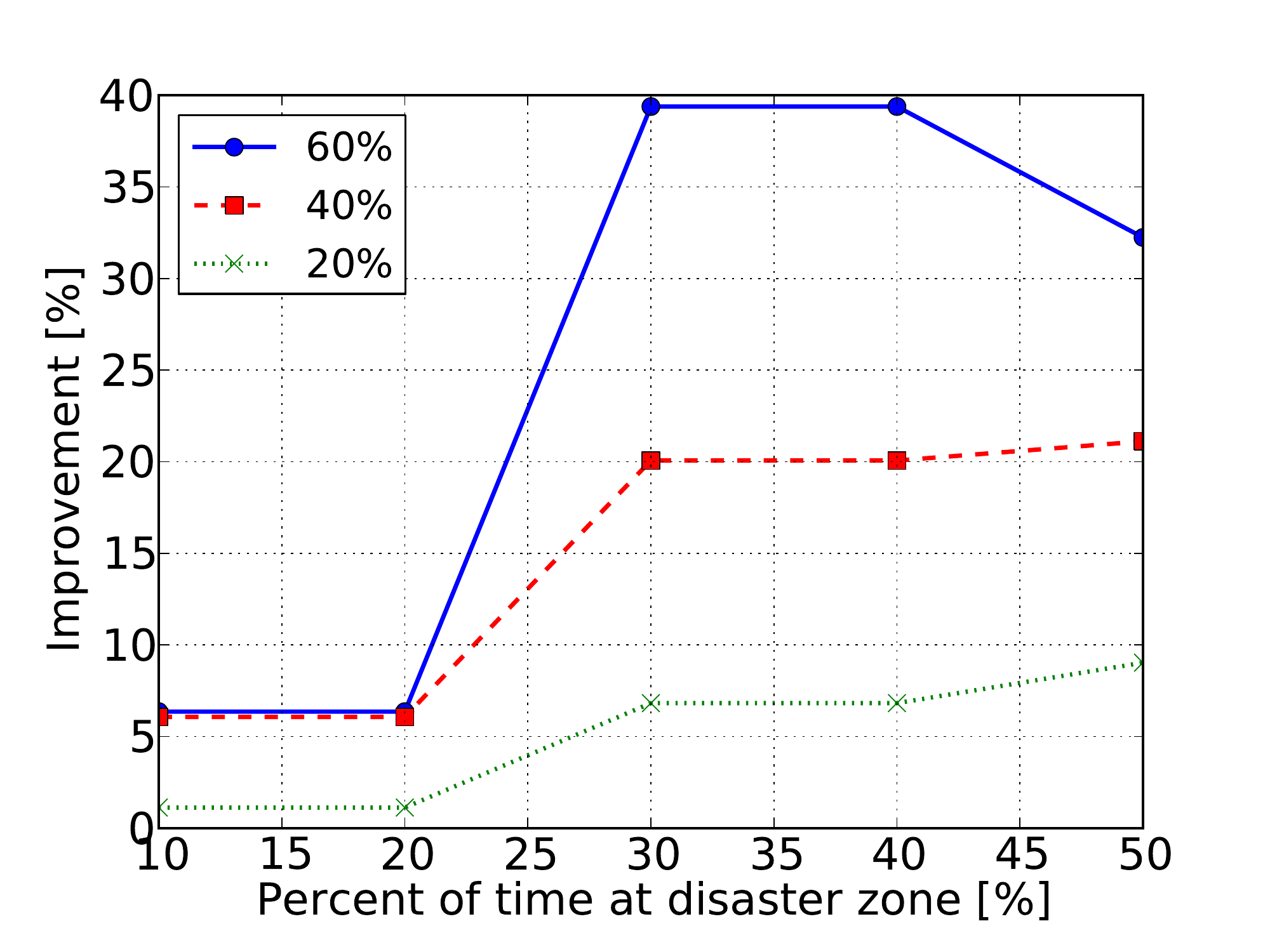}
    }
    \hspace{-7mm}
    \subfloat[Drops]{
    \label{fig:drops-vs-time}
	\includegraphics[width=0.34\textwidth]{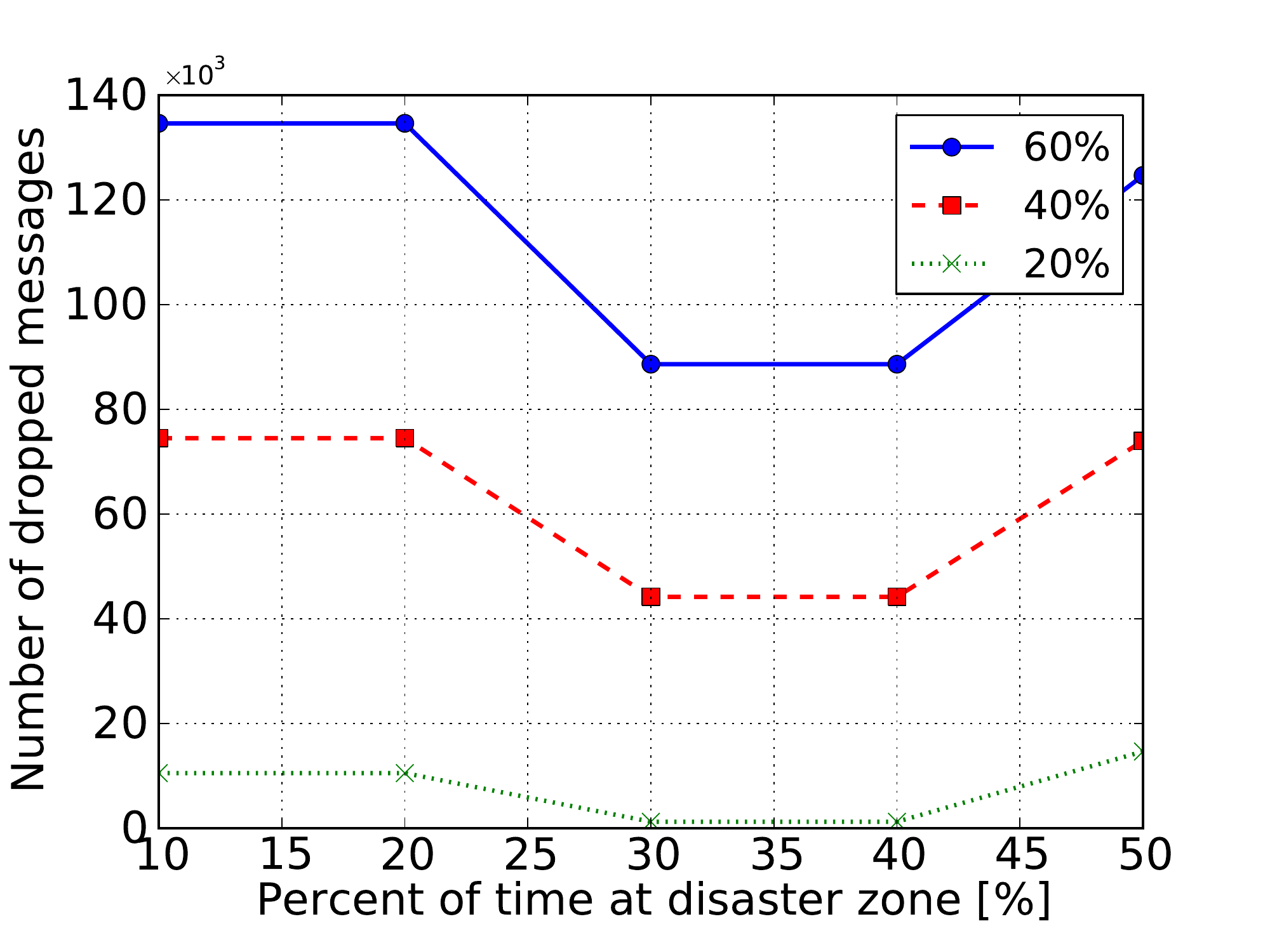}
    }
    \tightcaption{Simulation results, showing (a) an illustrative 5 hour run with 30\% redundancy, (b) percent of improvement in unique message delivery over non-RE for different $R_{sim}$ values, and (c) total number of dropped messages for different $R_{sim}$ values}
    \label{fig:visits}
\end{figure*}

When two devices are in contact range they exchange all the messages that
they have and the other party does not have, in a FIFO order.  We implemented
CARE into the ONE simulator, and created a CARE epidemic router. The CARE
routers exchange messages using the normal epidemic flooding, except that when
it receives a message which is similar to a message it already has in its
buffer, it does not accept the new message (for simplicity, we  implement a
drop policy for redundant content without de-prioritization).

A CARE router in the simulator receives   three parameters -- buffer size, FP
rate (FPR) and FN rate (FNR).  When FP$>0$, and a received
message is not similar to any message in the buffer, the router mistakenly
identifies the messages as similar with probability FP, causing the new message
to be dropped. Similarly, When FN$>0$, and a received message is similar to one
of the messages in its buffer, the router mistakenly identifies the messages as
non similar with probability FN, which results in occupying more space in its
buffer and sending both of them.

\ignore{
Under the assumption that all CARE routers use the same content-aware similarity detection technique, it is essential to ensure that
once a faulty decision regarding similarity is made, all CARE routers will make the exact same decision in the
future. To achieve this all decisions are stored globally, and when a CARE router needs to take
a decision, it first checks if this decision was already taken, and uses it in case it was.
}

\subsection{Results}
In this section we seek to explore the range of network parameters in which CARE brings benefits. In order to understand the interesting operating region for CARE, we first describe two extreme cases. The overall capacity of our system to deliver messages is defined by numerous parameters, including the message generate rate, the buffers, the contact opportunities, the bandwidth, the data exchange among nodes, and so on. At one extreme, if the aggregate capacity of the
system is much higher than the offered load, then it is capable of
storing and transmitting all the messages that
are created, and hence there would be no need for CARE capabilities. At the other extreme, when the system capacity is extremely
low, there is not enough storage and bandwidth to carry even the unique information to its destination, thus
CARE brings little or no added value. We study operating regimes in between these two extremes and focus on the impact to performance of the level of redundancy in the data, and the contact opportunities with the rescue vehicle. We sometimes compare our system to one without CARE, called non-RE, in which messages are served (and dropped when necessary) in FIFO manner.
%

We set the buffer size of the people devices,
$B_{people}$ to be able to include exactly all the unique messages that are generated. Recall that
each message is 300KB, and $G$ is the interval between messages, the buffer of each person has a size of:
$B_{people} = 300KB \cdot (1-R_{sim}) \cdot T/G$.  We selected this approach to buffer sizing as an attempt to run experiments in operating regions where the overall system capacity is not in either of the two extremes described above. (However, the buffer size is not the only parameter affecting whether or not we end up in the extremes.) The rescue vehicle has a fixed 1GB buffer.

\ignore{
As an illustrative example of a single simulation run with the above parameters, 30\% redundancy and optimal CARE, consider \figref{fig:sim-run}, which shows the number of unique messages
that reach the communication gateway with optimal CARE and without CARE over the duration of the simulation run.
Three exchanges of messages between the rescue vehicle and the communication gateway are visible. During
the first two, the offered load (i.e., total number of messages in the system) is small enough to be completely stored and delivered with and without CARE. However, as more messages are generated and duplicated with
epidemic routing, the buffers in the people devices get filled and messages are dropped. In the non-RE setting,
both unique and redundant messages get dropped, whereas in the CARE setting, only redundant messages get dropped, since each buffer is capable of holding all unique messages, and there are zero FN mistakes. Therefore, in the
final message exchange between the rescue vehicle and the communication gateway, there are roughly 130 more unique messages that manage to get through when CARE is used, which are roughly 17\% of the total messages generated.
}

First, we illustrate the impact of CARE on the delivery of {\em unique} messages. As an example, we ran an experiment with $T=5$ hours of simulation time, and  $R_{sim}=30\%$. Fig.~\figref{fig:sim-run} shows the number of unique messages
that reach the communication gateway with and without CARE over the duration of the simulation run.
The stairs in the plot are exchanges of messages between the rescue vehicle and the communication gateway. During
the first few, the offered load (i.e., total number of messages in the system) is small enough to be completely stored and delivered both with and without CARE. However, as more messages are generated and duplicated with
epidemic routing, the buffers fill up and messages are dropped. In the non-RE setting,
both unique and redundant messages get dropped, whereas in the CARE setting, only redundant messages get dropped since each buffer is capable of holding all unique messages and there are zero FN mistakes. In this example, we see 3 exchanges (at times 130, 180 and 260 minutes)
where the rescue vehicle is able to deliver a larger number of unique messages with CARE than without it. In the final exchange between the rescue vehicle and the communication gateway, there are roughly 200 more unique messages that manage to get through when CARE is used, which are roughly 17\% of the total messages generated.

\ignore{
\begin{figure}[h]
\begin{center}
\includegraphics[width=0.5\textwidth]{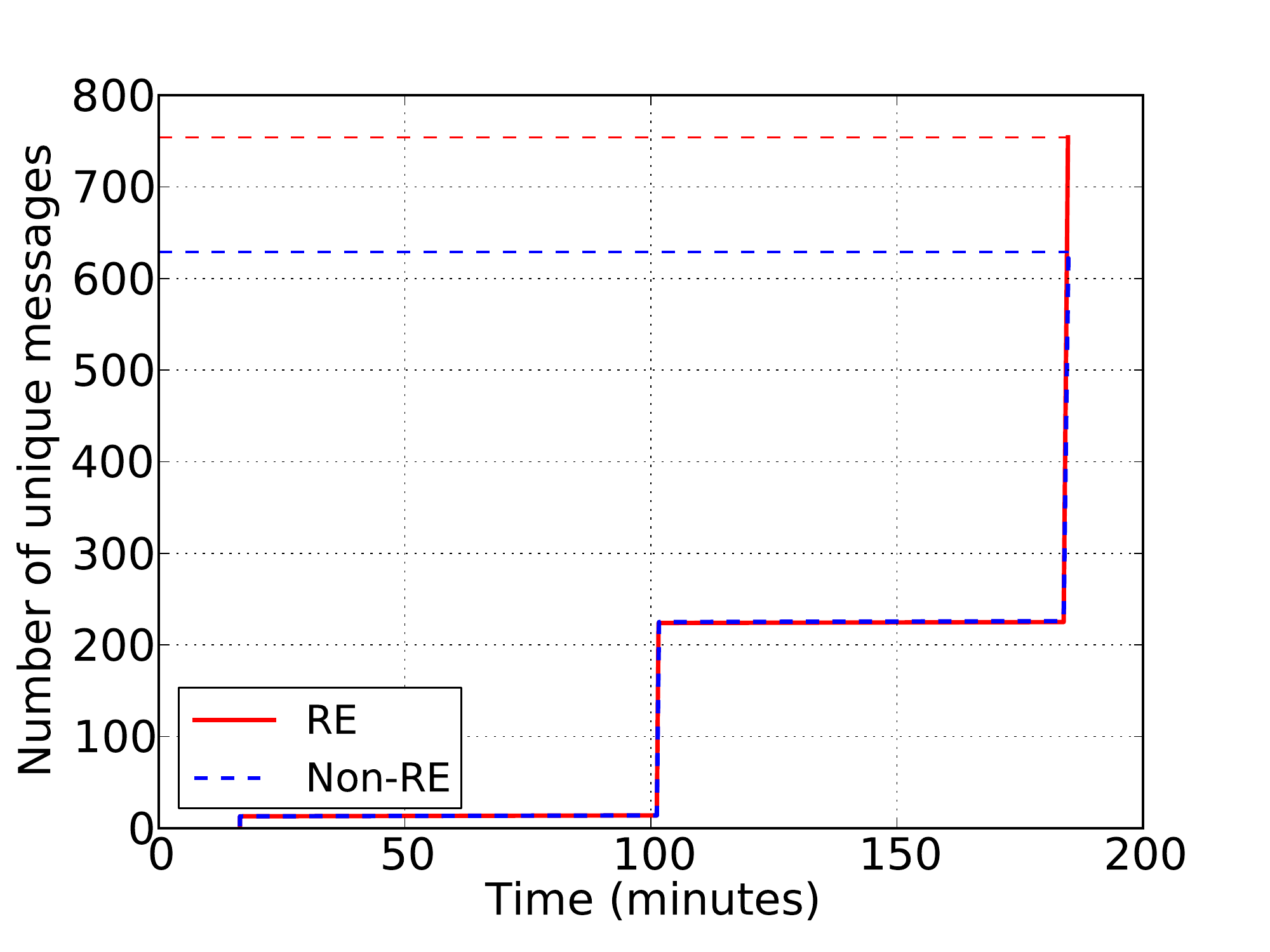}
\end{center}
\tightcaption{Number of unique messages across simulation time line}
\label{fig:sim-run}
\end{figure}
}

In order to quantify the improvement that CARE brings in delivery of unique information to the outside world we define the following.
Let $N^{CARE}_i$ be the total number of messages (including redundant messages) that reach the destination
gateway for a specific simulation run $i$ with CARE, and $U^{CARE}_i$ be the total number of unique messages (without redundant messages). Similarly, let $N^{Non-RE}_i$ and $U^{Non-RE}_i$ be the total number
of messages and the number of unique of messages respectively, in an identical but non-RE simulation setting.
The improvement is then given by:
$$\%Improvment = 100\cdot\frac{U^{CARE}_i-U^{Non-RE}_i}{U^{Non-RE}_i}$$

\figref{fig:imp-vs-time} plots the improvement for both various redundancy levels, and in terms of the percent of time that the rescue vehicle
spent at the disaster zone. We vary the percent of time
by changing the value of $Pr_{disaster}$. First we observe that the improvement increases as the redundancy in the traffic increases, implying that the more redundancy there is, the more benefits CARE brings. The performance is also a clearly function of how the the rescue vehicle partitions its time between the disaster zone and the outside world. When the time in the disaster zone is too low, the improvement is lower than 7\%, mainly
since there are not enough opportunities to convey information from the people in the disaster zone to the outside gateway. When the time in the disaster zone reaches 30\% to 40\% we seem to be in a good operating region as there are enough contact opportunities for smart dropping to take advantage of. In this region, we see an improvement ranging from 20\% when $R_{sim}=40\%$ to 40\% with $R_{sim}=60\%$.

We can see the effect of smart dropping via CARE in \figref{fig:drops-vs-time},
which shows the total number of message drops, i.e., drops that occur in the
buffers of all devices, when FIFO dropping is used. Looking at Figures 4b and
4c, it becomes clear that the improvement occurs because the drops under FIFO
are poor, whereas under CARE unique information is not lost. The number of
drops goes down as the rescue vehicle is spending enough time in the disaster
zone, and also has sufficient opportunities to convey the collected messages to
the outside.

\begin{figure}[th]
\centering
    \subfloat[]{
	\label{fig:latency-cdf}
	\includegraphics[width=0.37\textwidth]{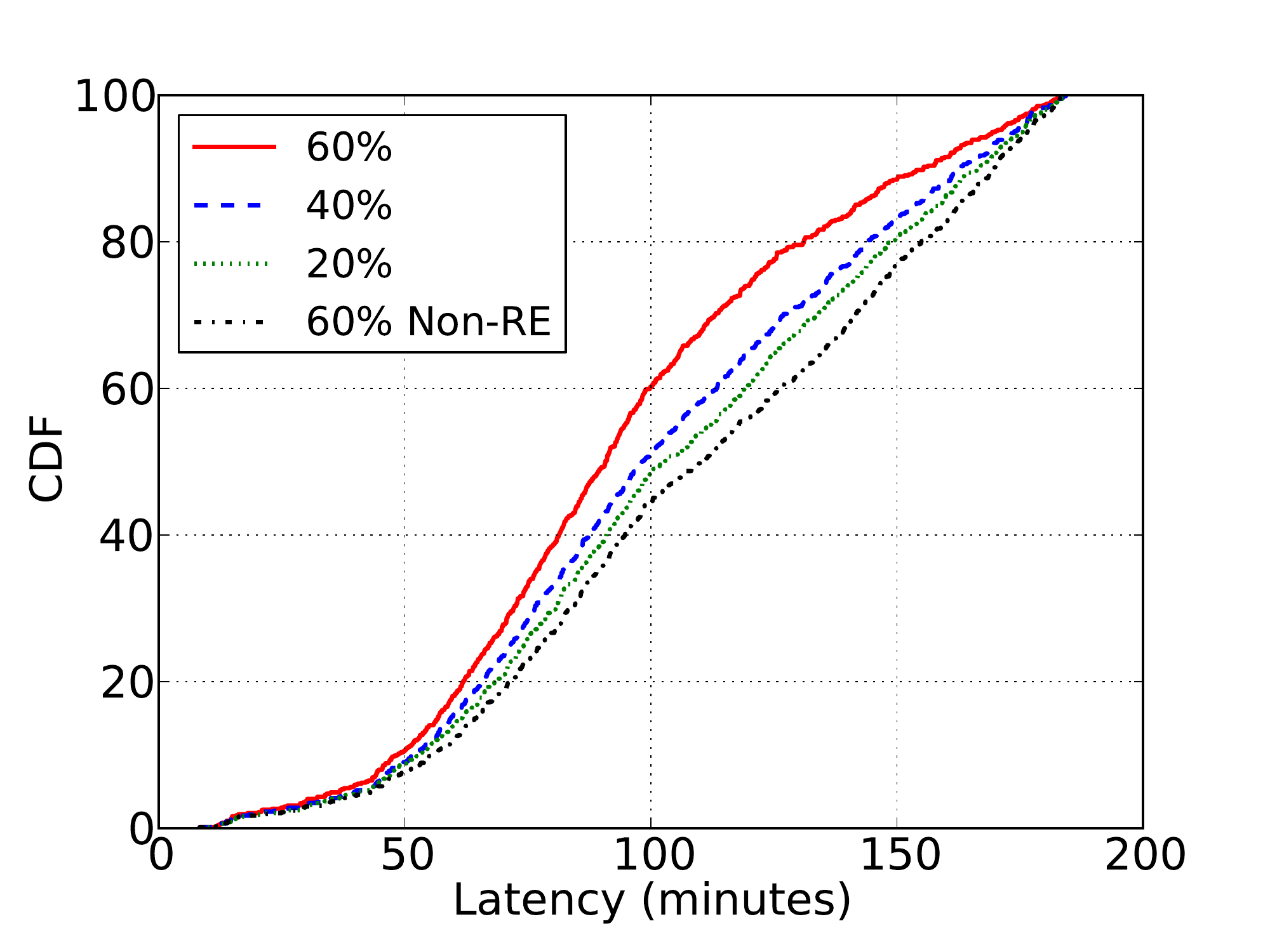}
    }
    \hspace{-7mm}
    \subfloat[]{
    \label{fig:latency-fps-cdf}
	\includegraphics[width=0.37\textwidth]{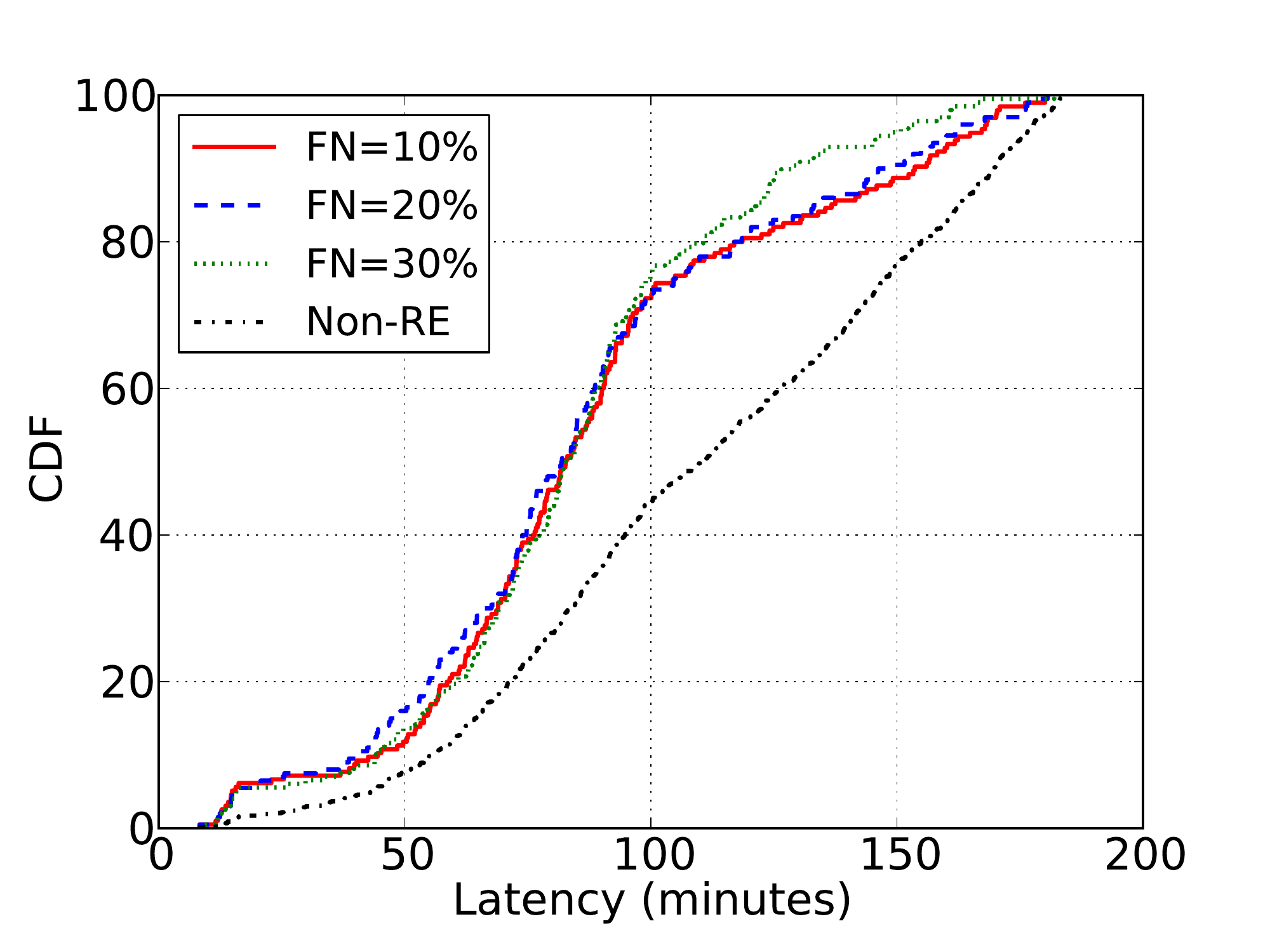}
    }
    \tightcaption{CDFs of message delivery statistics, showing (a) delivery time using varying redundancy levels, (b) delivery time of pipeline CARE. Plot (b) uses 40\% redundancy and FP rate of 1\%}
    \label{fig:latency}
\end{figure}

The reason the number of drops
increases with redundancy even in the non-RE setting is that
the buffer size $B_{people}$ is targeted to hold only unique messages, enabling
zero FIFO-related drops in all CARE settings. The more redundancy in the dataset, the smaller the buffer and
more drops are witnessed in non-RE settings.

Next we examine the impact of CARE on message delivery time (latency), i.e.,
the time it takes a message to reach the communication gateway.
\figref{fig:latency-cdf} shows an empirical CDF of message latency, for
different levels of redundancy.  In the case with 60\% data redundancy, then we
see that CARE can deliver the messages with significantly less latency than a
non-RE system. With CARE, 60\% of the messages will be delivered in under 100
minutes, while without CARE, those messages will require 140 minutes. Roughly
50\% of the messages will see almost 30 minutes improvement in median latency
(over 17\% of the maximal latency). This occurs because unique messages in a
FIFO system will get stuck in buffers behind redundant messages, whereas this
does not happen with CARE. We also see that as the redundancy increases, CARE
brings increasing benefits, as it reduces the overall latency of messages. Note
that even when running non-RE with the largest buffer size, which is obtained
using 20\% redundancy, its delivery time was smaller than all CARE simulations.
%

We note that in the above figures, we considered a version of CARE in which the image similarity detection is optimal and does not make mistakes. We considered this scenario only in order to isolate the effects of message redundancy, and rescue vehicle contact opportunities. This is important because there are so many parameters that ultimately affect the timely delivery of {\em unique} messages, and it is challenging to understand the impact of each. Next we considered non-optimal CARE based upon today's more realistic computer vision methods (although we are convinced that the performance of such algorithms will improve in the near future). \figref{fig:latency-fps-cdf} explores the impact of mistakes in image similarity detection on the delivery time, using a 4 hour simulation, 40\%
redundancy and a fixed FP rate of 1\%. Additionally, for these simulations we also made the buffer of the rescue vehicle
to be exactly the same size of the buffer of the people's devices, making the rescue vehicle capable of
storing exactly all the unique messages in the system.
The figure shows that the FN rate has little impact on the message delivery
times, and for all FN rates, CARE results in consistently better deliver times than non-RE. Although FN mistakes
increase the number of messages that traverse the network, the overall FN rate is still relatively small,
resulting in much less messages, which again reduces the load on the buffers.
%

\ignore{
\begin{figure*}[t]
\begin{center}
\begin{minipage}[t]{3.3in}
\centering
\includegraphics[width=\linewidth]{./figures/latency_40_disaster_40.pdf}
\tightcaption{Latency of delivering unique messages with 50\% redundancy in data and 40\% of the time in the disaster zone}
\label{fig:latency-cdf}
\end{minipage}\hspace{1.0em}
\begin{minipage}[t]{.48\textwidth}
\centering
\includegraphics[width=\linewidth]{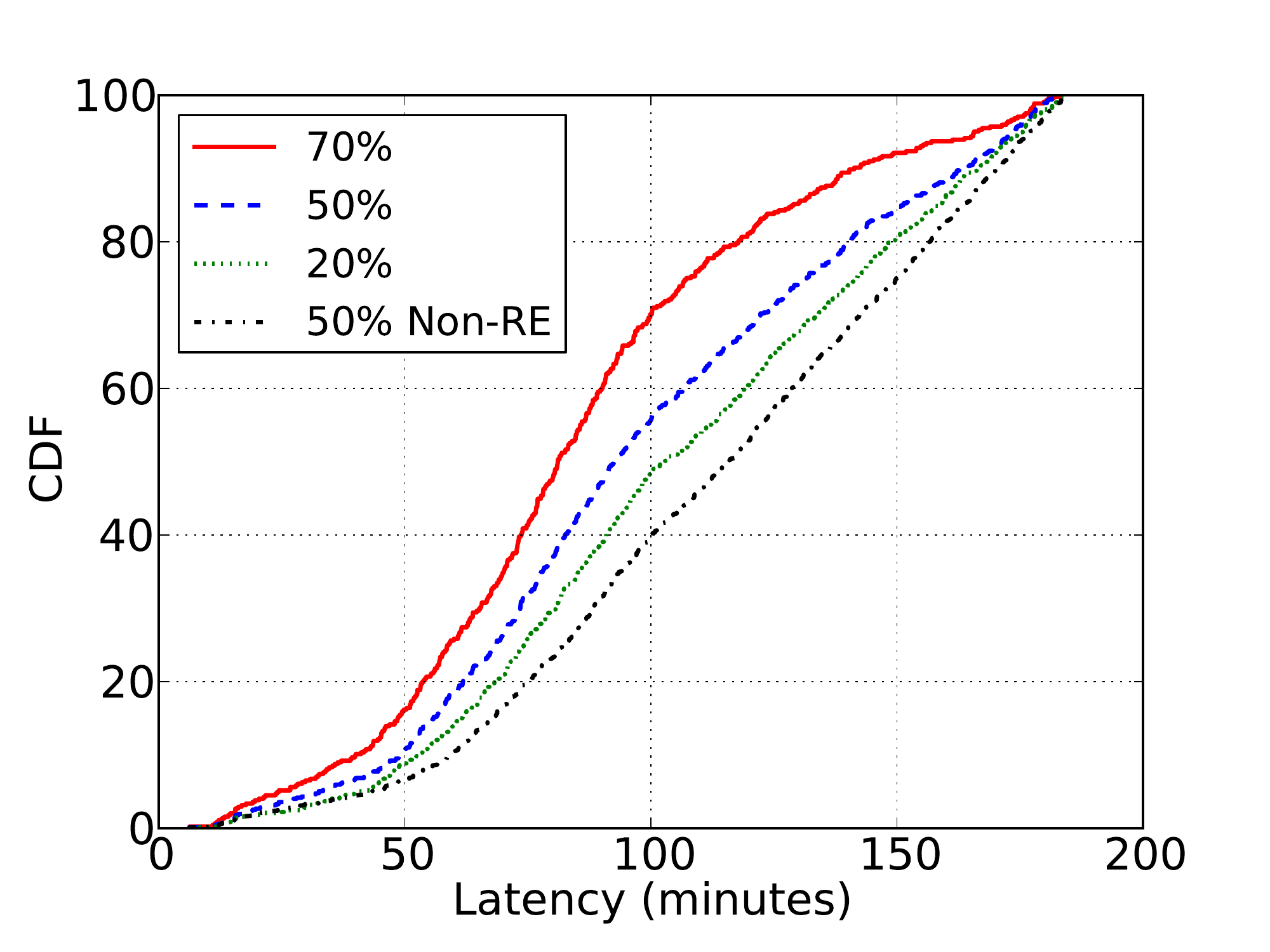}
\tightcaption{Latency of delivering unique messages in terms of redundancy with 40\% of the time in the disaster zone}
\label{fig:latency-redundancy-cdf}
\end{minipage}
\end{center}
\vspace{-1em}
\end{figure*}
}

\ignore{
[[\figref{fig:msgs-vs-redundancy} -- IT DOESN'T GIVE US ANYTHING UNTIL WE ADD RE MISTAKES]]

\begin{figure*}[tb]
\begin{center}
\begin{minipage}[t]{3.3in}
\centering
\includegraphics[width=\linewidth]{./figures/unique_over_time_30_5hour_1M.pdf}
\tightcaption{Time series plot of unique messages delivered}
\label{fig:delay}
\end{minipage}\hspace{1.0em}
\begin{minipage}[t]{.48\textwidth}
\centering
\includegraphics[width=\linewidth]{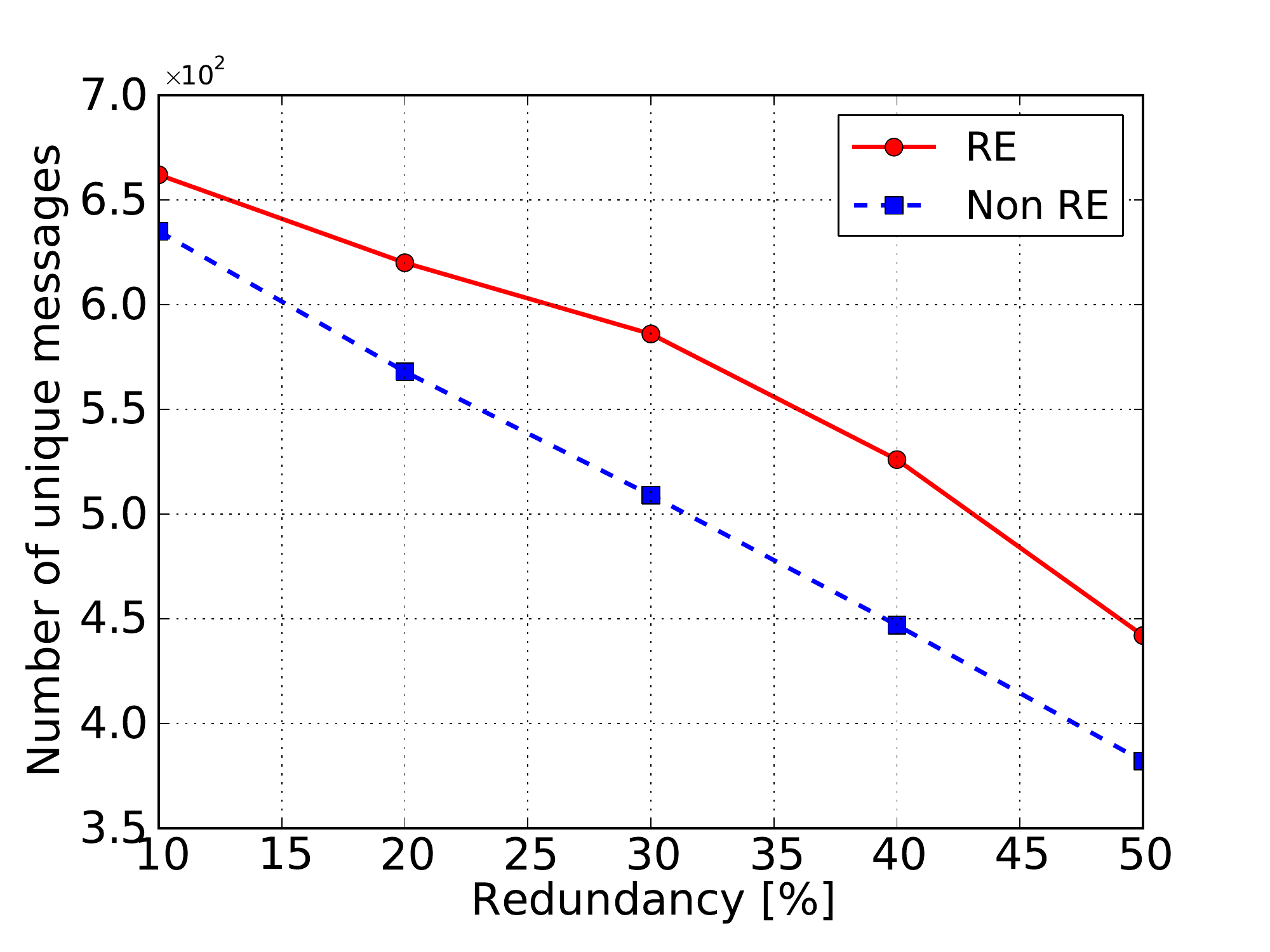}
\tightcaption{Delivery of unique messages as a function of the message redundancy}
\label{fig:redundancy-visits}
\end{minipage}
\end{center}
\vspace{-1em}
\end{figure*}
}

\subsection{Summary of Key Results}
To summarize, our evaluation shows that:

\begin{packeditemize}

\item CARE delivers roughly 20\% more unique information than a system without CARE.

\item The more redundancy there is, the greater  benefits CARE brings. For
example, we saw 20 to 40\% improvement, for redundancy ranging from 40 to 60\%
(note that our SDfire dataset falls in this range).

\item CARE not only delivers more unique information, but also does so in a
more timely fashion. In  our scenarios, we saw latencies reduced by 33\%.

\item Finally, CARE provides significant benefits  even with imperfect image
similarity detection, and a  conservative pipeline (i.e. don't drop messages
until they  passed numerous similarity checks). Even with false
negative rates of 10, 20 or 30\%, CARE
 still delivers more unique messages with lower latency than a non-RE system.

\end{packeditemize}

\section{Related Work}

PhotoNet \cite{photonet} presents a context-based redundancy elimination routing scheme that is motivated by the same scenario described in this paper, however, their work differs in key ways from ours. First, their method for assessing similarity in images is entirely different than ours. PhotoNet heavily relies on geo-tagging and photos within a short distance of each other are considered similar. To differentiate images from the same area, PhotoNet assesses similarity using color histograms. Although this approach has advantages computationally, it won't disguish critical differences for disaster relief. Consider as examples i) one person taking 2 photos in the same place but turned 180 degrees in the second photo relative to the first. Although the color tones in an area plagued by fires or hurricanes might be overall similar, the objects in the images will be different;  ii)  a burning house with no person in front is different than the same burning house with a policeman in front; the latter indicates that help is nearby and should be considered more important. The kinds of computer vision algorithms that we employ are able to do object extraction and thus provide fundamentally different options for assessing the notion of image similarity or importance.   Moreover,  even though most mobile phones nowadays have geo-tagged functionality, without the presence of cellular network this functionality relies on GPS, which consumes large amounts of battery power, thus will likely be turned off in a disaster scenario.  Second, the work in \cite{photonet} does not motivate why redundancy occurs, nor evaluate their protocol on real disaster data. In our work, we demonstrate the existance of redundacy in disaster photo data, and we believe that our work is the first to evaluate this approach on real data. A third key difference is the movement model. The authors proposed a model that assumes repetitive movement of people and rescue teams. This model increases the opportunities for redundancy elimination, which benefits the results of the simulation. We take a more conservative approach, and assume a random movement with only the rescue vehicle having a partially repetitive movement. We also explore a range of parameter settings to study under which operating conditions an approach like CARE brings benefits.

In \cite{allman} the author studies a similar scenario as ours in which a victim in a disaster zone is only allowed to send a limited number of small messages outside the disaster zone. The study focuses on how to set up a system in advance to allow a pre-specified set of people to receive such messages, and how to make the message upload procedure secure from malicious impersonation. There has been a large body of work on wireless protocols for emergencies, however that area of research is completely distinct from the discussion here that assumes that once a contact opportunity occurs the underlying wireless protocol (WiFi in this case) will work.

In the last few years, a number of techniques have been developed to identify
and remove strings of bytes that are repeated across network packets (e.g.,
\cite{RE-sigcomm08}). These techniques for traffic redundancy elimination are
content agnostic, and because they look for repeating byte patterns, these
techniques have been primarily applied to network traffic flows and files.
Because our focus is on a different media type, namely photos, content-agnostic
techniques offer little potential to further reduce the size of such content
because it is already compressed (e.g. JPEG) or because small differences in
photos (e.g. even illumination) will generate different underlying bit-patterns
of the digitized image. We demonstrated this in Sec. ~\ref{sec:comparison}. For
photos and video alike, content-aware techniques are needed for redundancy
identification. Previous efforts \cite{RE-sigcomm08} have argued for
integrating the traffic agnostic redundancy elimination methods into network
infrastructure and protocols. Our paper pushes this notion even further by
advocating for incorporating content-aware RE into network infrastructure to
handle redundancy of a broader set of media types.

In \cite{iscope} the authors design a system called iScope for personal image
management on mobile devices. They employ similar techniques such as SIFT to
enable content-based image search, both on a person's own device and across a
set of devices. Their goal is to enable rapid image search while designing to
meet power constraints. While \cite{iscope} employs similar image similarity
detection methods, both their application setting and their constraints are
distinct from ours.


\section{Conclusion and Future Work}

In this paper, we explored the idea of using content-aware redundancy
elimination  in the forwarding decision of the networking layer. We proposed
CARE, an architecture that combines such traffic reduction with existing DTN
protocols. This system is suited for emerging situational awareness applications 
 that empower ordinary citizens during disaster events. We believe this is an
interesting emerging area because the use of the Web to provide SA during
disasters is a trend that will continue to gain momentum -- especially since
the Internet has already proven effective in disaster response. There is a
great deal of effort in developing components of the ecosystem for such
systems, including crisis response data formats \cite{google-data-formats},
visualization techniques \cite{mashups}, portals \cite{portal-iscram08}, and
more. Many of the networking and systems issues surrounding information
overload, congestion, power consumption, collaborative decision making, data
sharing, security and privacy have yet to be fully understood.

Our study shows that incorporating solutions from image processing into the
forwarding path has a number of benefits. In the face of congestion, we enable
more unique information to escape from disaster zones when contact
opportunities are sporadic and unpredictable. We illustrated a disaster
scenario in which up to 40\% improvement in the delivery of unique information
can be attained using CARE.  We also showed that CARE decreases the latency of
non-redundant messages, compared to a system without redundancy elimination.
For many of the messages in our scenarios, the delivery time to the SA service
can be increased by 30 minutes to over an hour. 


We discuss that using image processing this way incurs a cost on people's
mobile devices; plus image similarity detection methods are not perfect and do
make mistakes. We designed a pipeline system that combines multiple algorithms
to balance the cost-accuracy tradeoff. We show that such a system can be used
either  with similar (FP,FN) performance as the best single method but at much
lower cost (roughly half), or it can be used with with better performance all
around (FP,FN,cost) but where the cost gains are in the 15\% range. We believe
that the algorithms from the image processing community will continue to
improve in accuracy, and thus such systems will become even higher performing
in the future.

We plan to expand our work  to other  disaster scenarios, and to
further evaluate the impact of our design on other constraints such as power
since energy is a critical resource in battery powered mobile devices. We
also plan to generalize our architecture to include other types of multimedia
data. In this paper, we focused on content-aware reduction for photos in
disasters where the incentive for content-aware reduction is motivated by
intermittent resources to transport the content, and because of the known
existence of semantic redundancy in the data. We believe however that our
approach could be made broader as there is potential to apply it to other
problem scenarios of information overload in networks, not related to
disasters. 


\newcommand{\para}[1]{\smallskip \noindent {\bf #1.}}
\para{Acknowledgements} The authors would like to thank Karl Grobl  and Team Rubicon for sharing their photos with us.


\bibliographystyle{abbrv}
\bibliography{references}

\end{document}